\documentclass[epj,nopacs]{svjour}
\pdfoutput=1
\usepackage[utf8]{inputenc}
\usepackage{caption,amsmath,amssymb,amsfonts,graphicx,cite,multirow,tikz,enumitem}
\DeclareCaptionType{Diagram}
\usepackage{pstricks,pdflscape,morefloats,algorithmicx,algpseudocode,slashed}
\usepackage[section]{algorithm}
\usepackage{booktabs}

\makeatother

\usepackage{color}
\usepackage{hyperref}

\usepackage{mathtools}

\newcommand\LR[1]{\multicolumn{1}{|c|}{#1}}

\preprint{CoEPP-MN-18-29, MCNET-18-29, KA-TP-34-2018, HERWIG-2018-04}

\title{Kinematic strangeness production in cluster hadronization}

\author{Cody B Duncan\inst{1,2}\and
Patrick Kirchgae\ss er\inst{2}}

\institute{School of Physics and Astronomy, Monash University, Clayton, VIC 3800, Australia \and Institute for Theoretical Physics, Karlsruhe Institute of Technology, 76128 Karlsruhe, Germany}

\date{\today}

\abstract{
We present a modification to the non-perturbative strangeness 
production mechanisms in the Monte-Carlo event generator Herwig in order 
to make the processes more dynamic and collective.
We compare the model to a series of observables for
soft physics at both LEP and LHC.}

\begin{document}

\maketitle

\section{Introduction}
The non-perturbative elements of simulating LHC events remain 
an active area of research in light of recent ALICE and 
CMS data\,\cite{ALICE:2017jyt,Khachatryan:2011tm}. Signs of strangeness 
enhancement and collective effects 
in high multiplicity events respectively have inspired several phenomenological 
models, ranging from interacting strings \cite{Bierlich:2014xba,Bierlich:2016vgw}, 
to relativistic 
hydrodynamics \cite{Pierog:2013ria}, to tweaks to the existing multiple parton 
interaction mechanisms \cite{Blok:2017pui} and colour reconnection 
\cite{Christiansen:2015yqa,Gieseke:2017clv} models. Monte Carlo event
generators \cite{Pierog:2013ria,Herwig7,Sjostrand:2014zea,Gleisberg:2008ta} provide a useful 
testing ground for these models.

Arguably the most successful models of hadronization which try to 
reproduce strangeness enhancement in 
high-multiplicity events are rooted in the physics of collectivity,
where the dense environment of high multiplicity events leads
to more complicated systems which interact with one another.
Heavy ion event generators typically prefer
a hydrodynamic viewpoint, where the quark-gluon plasma
acts as a perfect fluid, changing the dynamics of hadronization.
High-energy $pp$ event generators tend to use sophisticated 
iterations of the more conventional 
proton collision techniques, such as the DIPSY rope model where several
overlapping Lund strings \cite{ANDERSSON198331} 
combine into a higher-representation colour field, which
then may enhance strangeness production and may also shove each other
transversely outwards, mimicking the fluid behaviour of quark-gluon
plasma. Another model \cite{Fischer:2016zzs} has attempted to use a thermodynamics
inspired route to string fragmentation and was able to explain a harder 
transverse momentum spectrum for heavier particles.

Herwig \cite{Herwig7} has recently developed a new model for colour reconnection, 
where baryonic clusters were allowed to be produced in a geometric 
fashion \cite{Gieseke:2017clv}, in an attempt to explain the results of \cite{ALICE:2017jyt}.
The model was able to create heavier hadrons, and in particular more baryons,
but in order to better describe the data, the non-perturbative gluon splitting
mechanism was allowed to produce $s\bar{s}$ pairs as well as the default lighter species.
However, the production weight was simply set to a 
flat number, tuned to Minimum Bias events at the LHC.
In this paper, we will mainly focus on the fundamental mechanisms 
of strangeness production in cluster hadronization, namely the production 
rate of $s\bar{s}$ pairs during non-perturbative gluon splitting, 
cluster fission, and cluster decay. In doing so, we are
taking the first steps to a rework of strangeness 
production in the Herwig hadronization phase. A full model would also need to consider 
colour reconnection, since this rearranges the colour topology and thus 
the mass distribution inside an event, affecting the scaling 
that we are interested in studying.

In this study, we aim to introduce a simple dynamic model of strangeness production 
in Herwig, in which each non-perturbative production stage 
uses the kinematic information of the relevant surrounding colour-singlet 
system. After reviewing the current mechanisms of hadronization in 
Sec.\,\ref{sec:hadronizationmodel}, we perform two separate tunes to a number
of light strange meson observables for LEP and LHC Minimum Bias events in
Sec.\,\ref{sec:oldtune}. We show that the tuned current strangeness production
parameters are drastically different between the two collider types, and propose
a mass-based scaling for the relevant production weights in Sec.\,\ref{sec:newmodel},
comparing two different mass-like measures to scale the probability.
In Sec.\, \ref{sec:analysis}, we tune our new model and compare the results
with the old model in Herwig, as well as
perform a comparison to the default Lund string model in Pythia \cite{Sjostrand:2014zea}
with the Monash tune \cite{Skands:2014pea}.
We briefly summarize the work and possible future avenues for
research in Sec.\,\ref{sec:conclusion}.

\section{The Herwig Hadronization Model}
\label{sec:hadronizationmodel}
To accurately describe a full QCD event, one must be able to model the 
non-perturbative physics contributions, e.g. hadronization of individual quarks 
\& gluons from the parton shower
and the multiple parton interactions to form colour-singlet hadrons. 
\begin{figure}[t]
\centering
\includegraphics[width=0.49\textwidth]{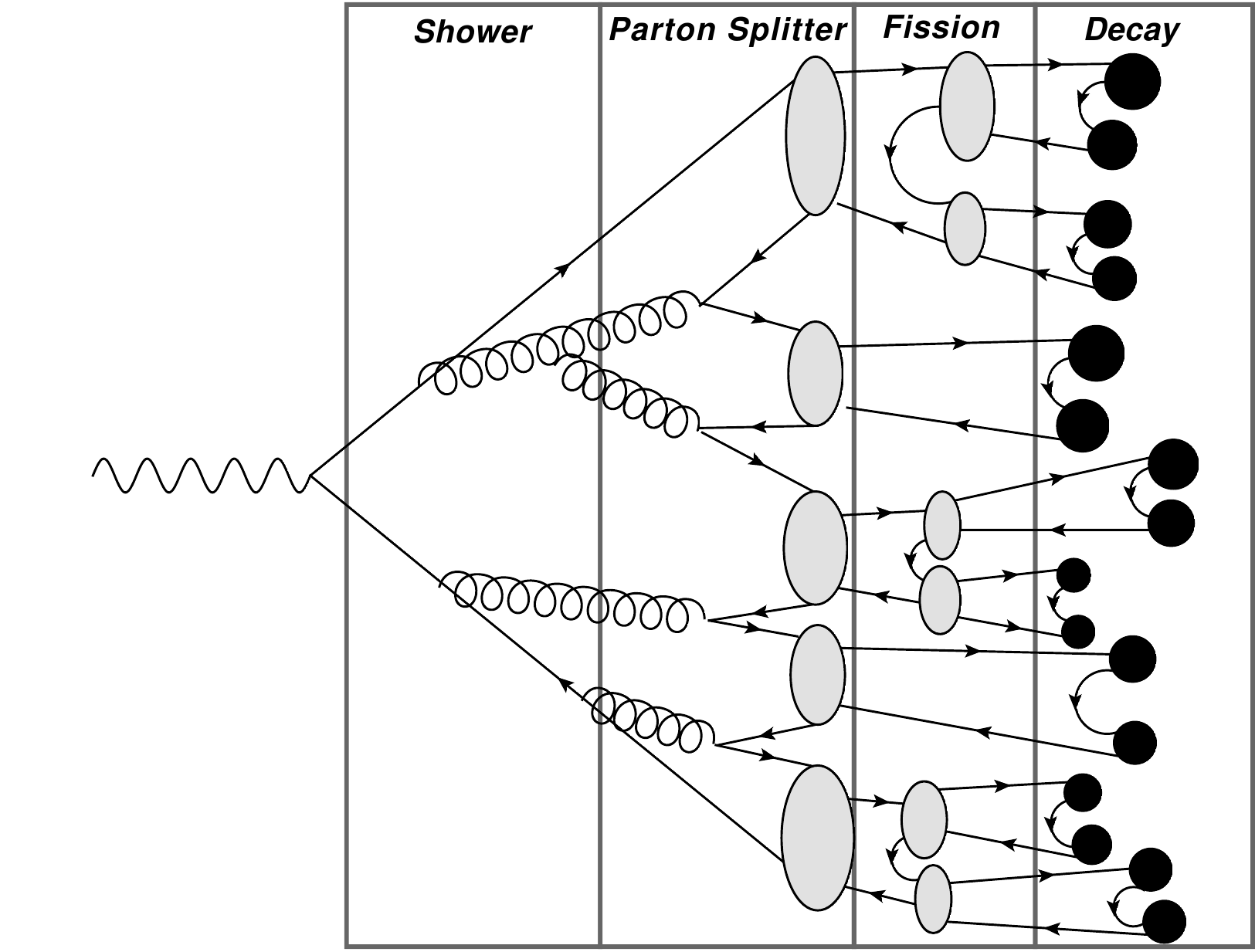}
\caption{Figure of a simplified event where we show the major stages of 
hadronization after the parton shower that can contribute to non-perturbative 
strangeness production. Grey ellipses are clusters, while black are hadrons.
}
\label{fig:eventgenerator}
\end{figure}

Fig. \ref{fig:eventgenerator} sketches a schematic event, focusing
on the final state. After generating a hard matrix element for the event, Herwig performs a 
parton shower, producing a number of soft and collinear partons. After the parton shower 
reaches $\mathcal{O}(1)$ GeV, the hadronization phase of simulation occurs.
In Herwig, the hadronization model is the cluster model \cite{WEBBER1984492}, 
based on the colour preconfinement \cite{AMATI197987} property 
from the angular-ordered parton 
shower. A cluster can be considered to be a highly primordial, excited colour-singlet 
$q\bar{q}$ pair. 

There are several parts to the hadronization model in Herwig, 
in the following algorithmic order:
\begin{itemize}
	\item Non-perturbative gluon splitting,
	\item Colour reconnection,
	\item Cluster fission,
	\item Cluster decay to hadron pairs,
	\item Unstable hadron decays.
\end{itemize}
In Fig. \ref{fig:eventgenerator}, we have omitted colour reconnection
since this step simply changes
the colour topology of the event, not the content of the clusters.
While modifying the colour reconnection algorithm would have a non-trivial
impact on the later stages of hadronization, namely cluster fission and decay,
it is outside the scope of this paper, but these correlations will 
be studied and addressed in future work. Since the scope of this project is 
mainly focused on light strange hadron production, we tune predominately 
to pion and kaon observables. We will also ignore unstable hadron
decays for the purposes of this paper.

The three other listed stages in hadronization are each allowed to contribute to the overall
strangeness in the event, since they each produce new $q\bar{q}$ pairs.
We briefly recall the details of each step as presented in depth in \cite{Herwig7}.

\subsection{Non-perturbative gluon splitting}
Once the parton shower ends, all gluons undergo a non-perturbative splitting
into $q\bar{q}$ pairs. The species of the pair is determined by a given weight,
e.g. in the tune from \cite{Gieseke:2017clv} the weights of up, down, 
and strange are 2:2:1. The default version of Herwig does not allow for strangeness
production at this step, only $u\bar{u}$ and $d\bar{d}$ pairs.
The only constraint on the gluon splitting
is that the gluon mass is at least twice the constituent mass of 
the species in question, and the gluons are split isotropically.

After all the gluons in an event have been split, nearest neighbours in
momentum space are most likely to be nearest neighbours in colour space
\cite{AMATI197987}, and clusters are formed from the momentum-space
neighbouring $q\bar{q}$ pairs, with a mass distribution decoupled from the hard
scattering process that created them.

\subsection{Cluster fission}
Exceptionally heavy clusters are allowed to fission into two lighter, less 
excited clusters if the mass $M$ of the original cluster satisfies the condition:
\begin{equation}
	M^p \geq q^p + (m_1+m_2)^p ,
\end{equation}
where $p$ and $q$ are parameters that control the fissioning rate criteria, 
and $m_{1,2}$ are the parton masses of the heavy cluster.
In Herwig, $p$ is given separate values for light quarks ($u,d,s$),
charm, and bottom. The light quark weights are further subdivided, and
strangeness is suppressed by a flat weight. $q$ has a similar divide between
the quark species.

After selecting clusters to fission, the cluster fissioner 
produces a $q\bar{q}$ pair from the light quarks with a fixed weight, 
distinct values for each flavour of
quark (bar top), and diquarks. Each parton from the pair 
go into a separate cluster, giving the new pair of clusters a mass distribution of:
\begin{equation}
	M_i = m_i + (M-m_i-m_q)R_i^{1/w} ,
	\label{eq:fission}
\end{equation}
where $w$ is the splitting parameter that controls the rate of splitting
for clusters containing different species of quarks.

\subsection{Cluster decay}
The last stage of cluster-based physics is at the cluster decay level, in which clusters
decay into excited hadrons. Given a cluster with constituents $q_1,\bar{q}_2$, the weight
for producing hadrons $h_a = q_1\bar{q}, h_b = q\bar{q}_2$, where $q$ denotes a quark
or diquark species, is given by:
\begin{equation}
	\mathcal{W}(h_a,h_b) = P_{q} w_as_aw_bs_bp^{*}_{a,b} ,
	\label{eq:decay}
\end{equation}
where $P_{q}$ is the production weight for the given quark or diquark species,
$w_i$ are the weights for the relevant hadron production, and $s_i$ are the
suppression factors for the corresponding hadrons. The final factor in the weight is the
two-body phase space factor that controls how readily the cluster can decay into the two
chosen hadrons.

\subsection{Herwig strangeness parameters}
The Herwig parameters that control non-perturbative\\ strangeness 
production are the gluon splitting weight - \mbox{\tt SplitPwtSquark}, 
and the cluster fission \& decay weight - \mbox{\tt PwtSquark}. 
In the original model, cluster fissioning and cluster decaying are 
controlled by the same parameter. The first step in our understanding 
of the different contributions is to disentangle cluster fission 
from cluster decay and introduce one additional parameter which 
controls the production of a $s\bar{s}$ pair during cluster fission 
- \mbox{\tt FissionPwtSquark}. The decay parameter remains the same.

\section{Tuning of the existing model}
\label{sec:oldtune}
In this section we tune the parameters for strangeness
production of the existing model first to LEP and then to LHC data.
Hadronization models are typically tuned to LEP data if they do not rely on
$pp$-specific event topology, e.g. multiple parton interactions
 and their effects on colour reconnection, since
LEP provides a clean QCD final state environment which imposes relatively strict
constraints on what one's hadronization model is allowed to do.

The tuning is achieved by using the Rivet and Professor frameworks for
Monte Carlo event generators\,\cite{Buckley:2009bj,Buckley:2010ar}.
In order to understand the overall effects of strangeness production 
on different stages of the event generation, we keep all other hadronization 
parameters that were previously tuned to LEP data at their default values \cite{Herwig7,Gieseke:2012ft}.
In the first tune (TUNE1), we only consider the effects of the 
parameters that are \textit{directly} 
responsible for strangeness production as explained in 
Sec. \ref{sec:hadronizationmodel}.

In a second tuning attempt (TUNE2), we introduce the new parameter for the
cluster fission stage. 
Tuning these 3 different parameters will allow us to study the phases 
of strangeness production during event generation and will shed light on 
the differences between LEP and LHC.  

We note that this section is an extended part of the introduction
to visualize and highlight the effects of the aforementioned different parameters
and to see at which stage non-perturbative strangeness production is preferred.


\subsection{LEP Tuning}
For the tuning to LEP data, the following observables from 
ALEPH \cite{Barate:1996fi,TheALEPHCollaboration2004}, DELPHI \cite{Abreu:1996na}, 
SLD \cite{Abe:2003iy} and PDG hadron multiplicities 
\cite{Amsler:2008zzb}, which represent a good description 
of event shapes and $\pi$, $K$ multiplicities, were used 
with equal weights:
\begin{itemize}
	\item Mean charged multiplicities for 
	      rapidities $|y|<1.0,\\|y|<1.5$ and $|y|<2.0$
	\item $K^0$ spectrum
	\item Mean $\pi^0$ multiplicty
	\item Mean $K_S+K_L$ multiplicity
	\item Mean $K^0$ multiplicity
	\item Mean $\pi^+/\pi^-$ multiplicty
	\item Mean $K^+K^-$ multiplicity
	\item Ratio (w.r.t $\pi^{\pm}$) of mean $K^{\pm}$ multiplicity
	\item Ratio (w.r.t $\pi^{\pm}$) of mean $K^{0}$ multiplicity
	\item $K^{\pm}$ scaled momentum
\end{itemize}
The resulting parameter values for the two different tunes 
are listed in Tab.\ref{table:table1}.

\begin{figure*}[th]
\centering
\includegraphics[width=0.49\textwidth]{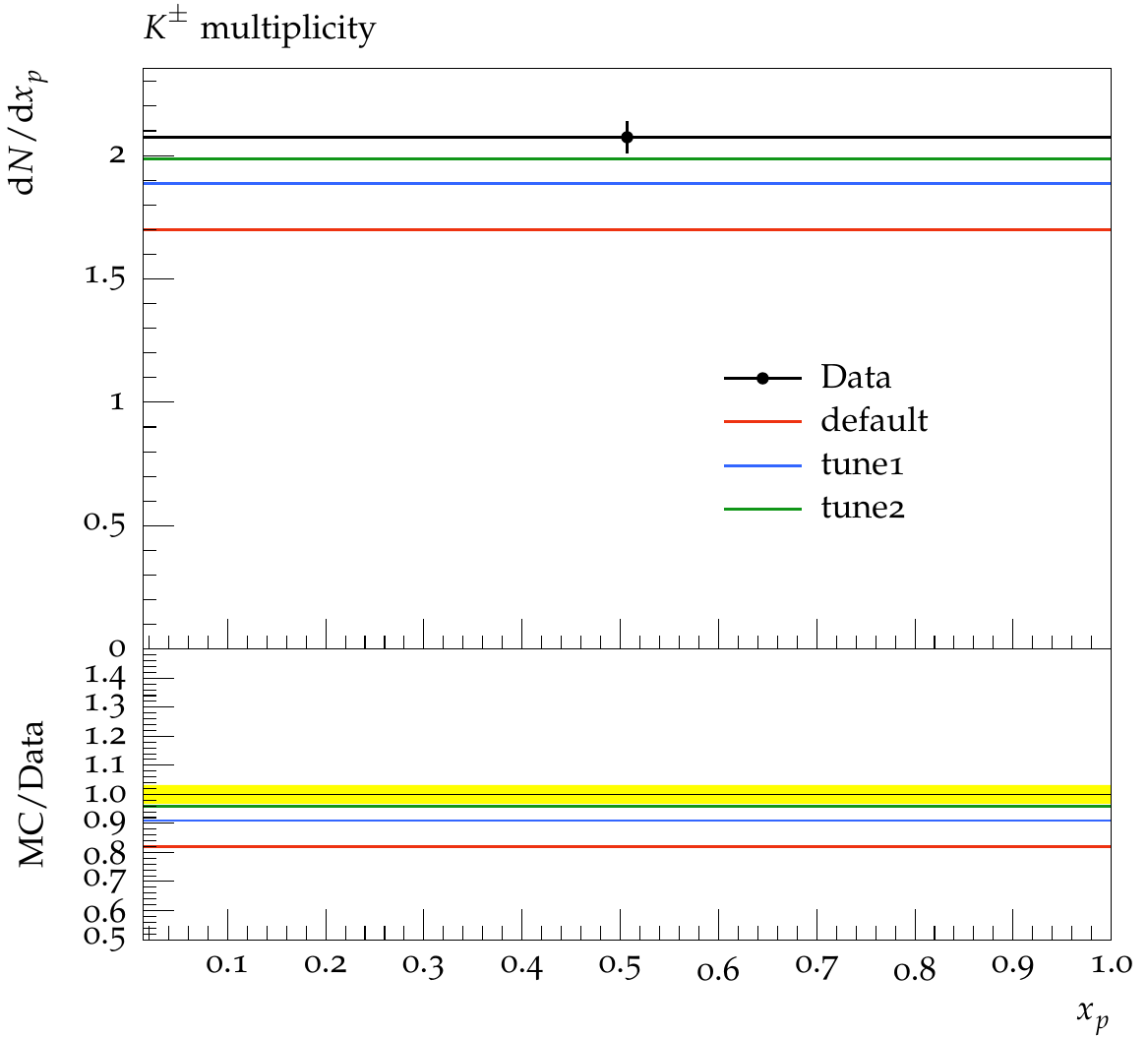}
\includegraphics[width=0.49\textwidth]{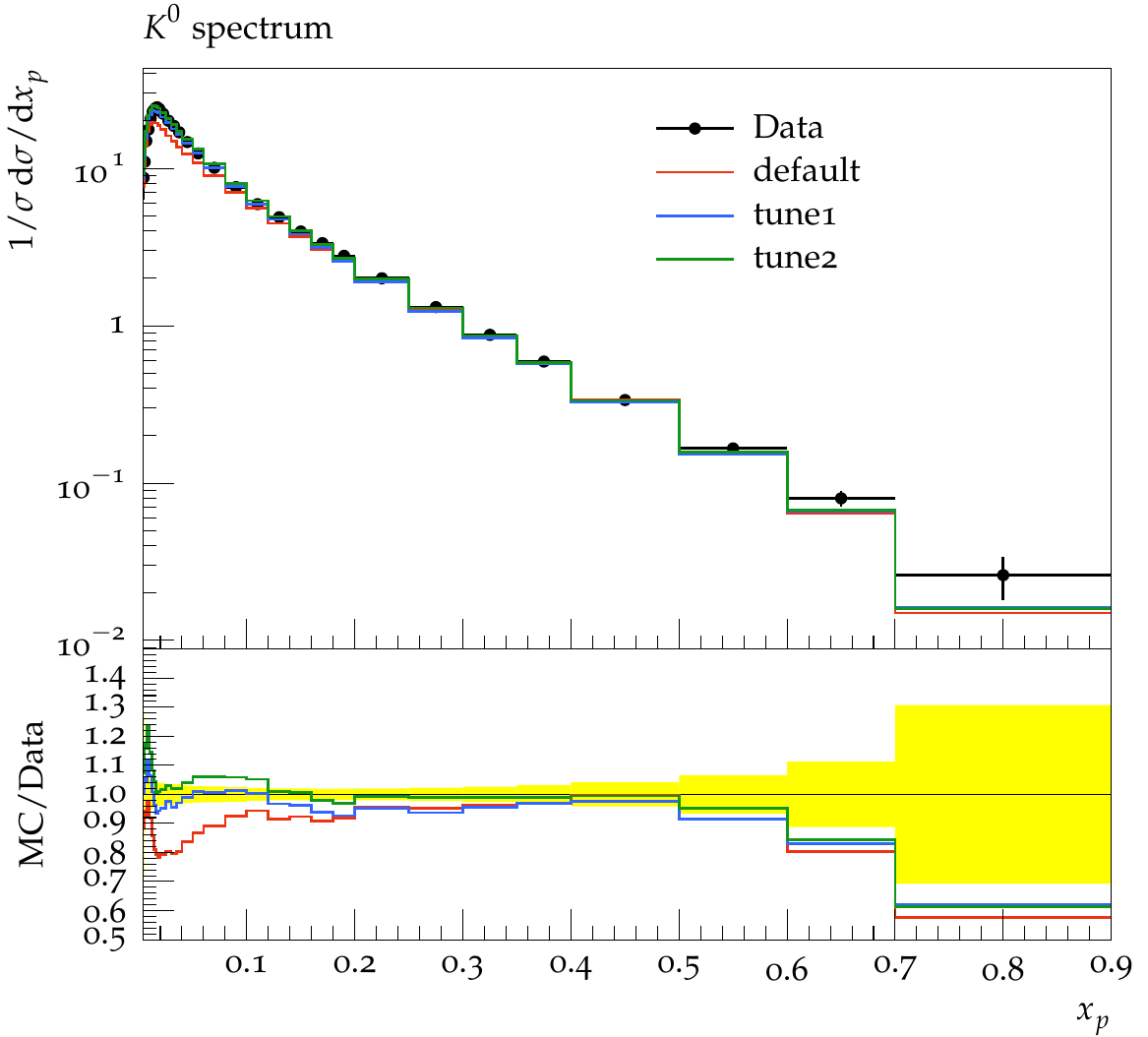}
\caption{ Measurement of $K^{\pm}$ multiplicities at SLD \cite{Abe:2003iy}
and $K^0$ spectrum as measured at ALEPH \cite{Barate:1996fi} for 
$\sqrt{s}=91.2\,\mathrm{GeV}$. We show a comparison between the default Herwig model
and our two different tunes.  
}
\label{fig:leptuning}
\end{figure*}

\begin{table*}[htpb]
\parbox{0.45\linewidth}{
\centering
\begin{tabular}{lccc}
    \textbf{LEP}  & Default & TUNE1 & TUNE2 \\
\midrule
Gluon Splitting  & -- & 0.24 & 0.19\\
\cline{2-3}
Cluster Fission &  \LR{0.66} &  \LR{0.53} & 0.69\\
Cluster Decay  & \LR{0.66}  & \LR{0.53} & 0.69\\
\cline{2-3}
\end{tabular}
\caption { Results of the parameter values for strangeness production at the different
stages of the event generation (LEP). In both default Herwig and TUNE1, 
cluster fission and decay have the same
parameter. In TUNE2, they are allowed to be different, but the tuning 
procedure returned equal values. 
In default Herwig, there is no $g\to s\bar{s}$ option.
  }
  
\label{table:table1}
}
\hfill
\parbox{0.45\linewidth}{
\centering
\begin{tabular}{lccc}
   \textbf{LHC}  & Default & TUNE1 & TUNE2 \\
\midrule
Gluon Splitting  & -- & 0.95 & 0.95\\
\cline{2-3}
Cluster Fission & \LR{0.66} &  \LR{0.05} & 0.02\\
Cluster Decay  & \LR{0.66}  & \LR{0.05} & 0.25\\
\cline{2-3}
\end{tabular}
\caption { Results of the parameter values for strangeness production at the different
stages of the event generation (LHC). In both default Herwig
 and TUNE1, cluster fission and decay have the same
parameter, while in TUNE2 they are allowed to be different. 
In default Herwig, there is no $g\to s\bar{s}$ option.
  }
\label{table:table2}
}
\end{table*}
While being able to describe all the considered LEP data 
on equally good footing, we improve the simulation 
of the observables which were considered in the tuning 
procedure. TUNE2 gives better agreement to the data,
at least with respect to the $K^{\pm}$ multiplicity,
highlighting the necessity to
disentangle the cluster fission and cluster decay parameters. 
The corresponding plots are shown in Fig.\,\ref{fig:leptuning}, 
where we compare the default version with our two new 
tunes.

\subsection{LHC Tuning}
For the tuning to LHC data, we solely focus on identified particle 
distributions which were measured at ALICE \cite{Adam:2015qaa} and CMS\,
\cite{Khachatryan:2011tm}. We limit the tuning to a center of mass energy 
of $\sqrt{s}=7\,\mathrm{TeV}$ due to the lack of suitable available 
Rivet analyses at higher energies. 
The following observables were considered in the tuning procedure with 
equal weights:
\begin{itemize}
	\item $K^++K^-$ yield in INEL pp collisions at 
	      $\sqrt{s}=7\,\mathrm{TeV}$ in $|y|<0.5$
	\item $K/\pi$ in INEL pp collisions at
              $\sqrt{s}=7\,\mathrm{TeV}$ in $|y|<0.5$
	\item $K_S^0$ rapidity distribution at $\sqrt{s}=7\,\mathrm{TeV}$
	\item $K_S^0$ transverse momentum distribution at $\sqrt{s}=7\,\mathrm{TeV}$
\end{itemize}
The resulting parameter values are shown in Tab.\,\ref{table:table2}.

\begin{figure*}[th]
\centering
\includegraphics[width=0.49\textwidth]{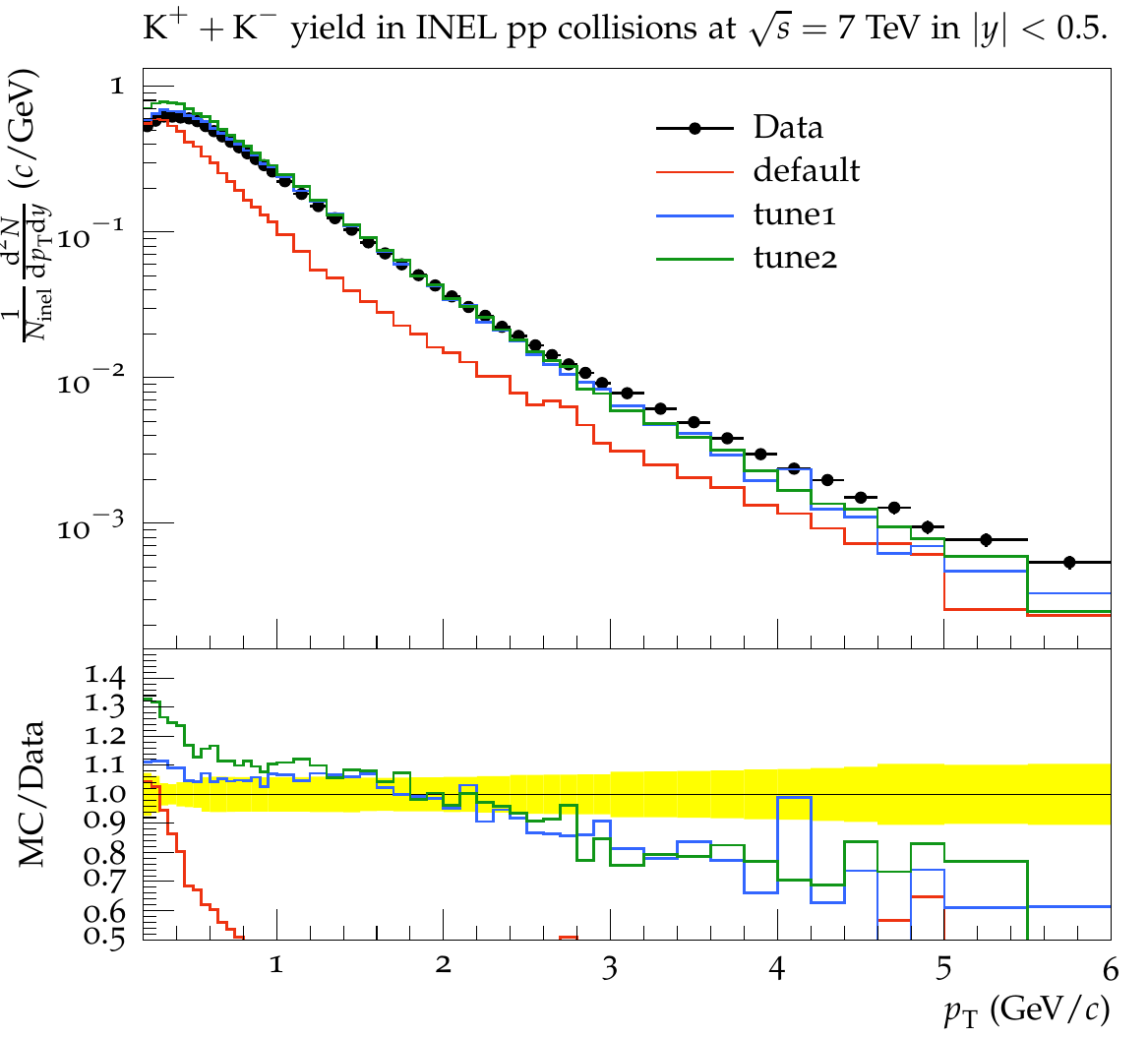}
\includegraphics[width=0.49\textwidth]{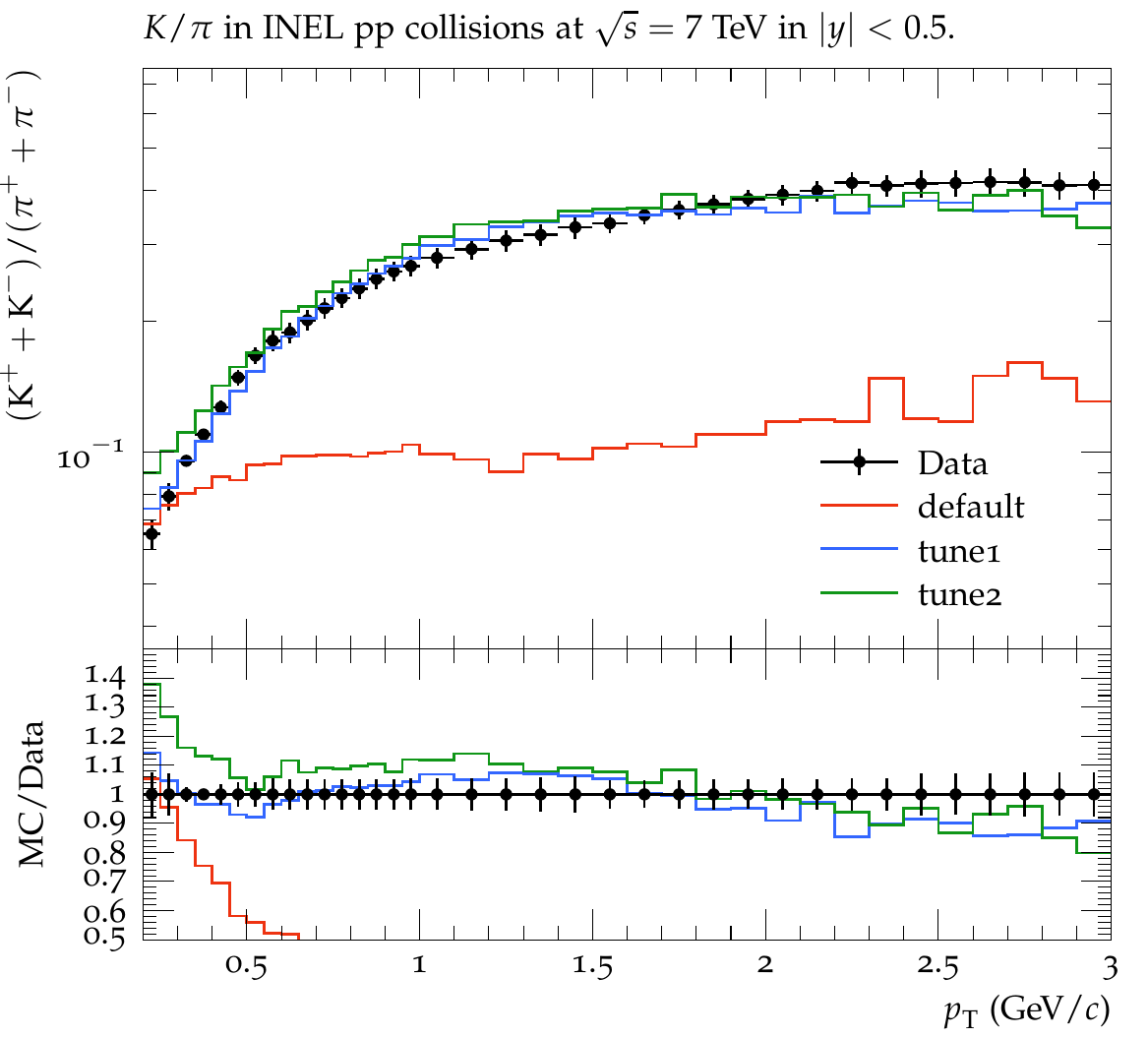}
\caption{Transverse momenta spectra for $K^++K^-$ and $K/\pi$ ratio as 
measured by ALICE\cite{Adam:2015qaa} at $\sqrt{s}=7\,\mathrm{TeV}$ in 
the central rapidity region. We show a comparison between the default Herwig
model and our different tunes.
}
\label{fig:lhctuning}
\end{figure*}

The outcome of the tuning procedure is shown for the $p_T$ distribution of $K^++K^-$ 
yields and the $K/\pi$ ratio in Fig.\,\ref{fig:lhctuning}.
Again the retuning of the default model with the incorporation of an additional
independent parameter at the cluster fission stage
improves the description of the considered observables significantly. 


\subsection{Summary}
The general approach in tuning a hadronization model is to tune the parameters to
LEP data and then assume it is able to describe LHC observables as well since 
hadronization is assumed to factorize and should not depend on the process involved.

The main difference between LEP and LHC is the denser hadronic environment
one encounters due to multiple parton interactions and therefore also the enhanced 
effect of colour reconnections on the distribution of final state particles.
Be that as it may, we believe that the probability to produce strangeness e.g at the stage of
non-perturbative gluon splitting should be a universal parameter 
and be independent of the process in question.

Since the data shows that clearly different parameter values are preferred
at LHC and LEP the approach to have a single valued probability is not suited
for the description of both LHC and LEP observables. It may capture the average effect
but it does not allow for fluctuations on an event-by-event basis. We tackle this problem 
by assuming that the rate at which strangeness is produced depends 
on the hadronic density of the immediate environment, which will be discussed 
in the next section. 


\section{Kinematic strangeness production}
\label{sec:newmodel}
As mentioned above, the various splitting probabilities and weights are 
flat numbers tuned to data, without any considerations for the topology 
of a given event. In order to have a more dynamic picture, where the 
splitting probabilities depend on the environment, 
we choose to scale the weights with respect to colour-singlet masses. 
The mass of a colour-singlet system 
at a given phase of hadronization scales the probability for 
strangeness production up or down,
depending on a characteristic mass scale for each step.

As a simple starting point for mass-based power scaling, we replace the flat weights
in each of the steps mentioned in Sec.\,\ref{sec:hadronizationmodel} 
with the following functional form:
\begin{equation}
	w_s(m)^2 = \exp\left(\frac{-m_0^2}{m^2}\right) ,
	\label{eq:scaling}
\end{equation}
where $m_{0}^2$ is the characteristic mass scale for each phase, 
and $m^2$ is the total invariant mass of the relevant colour-singlet system. In this work,
we will introduce another mass-based measure which replaces $m^2$ in the denominator of
Eq. \ref{eq:scaling}: the threshold production measure, $\lambda$. 
We discuss the difference in the
two approaches in Sec.\,\ref{sec:lambda}. For now, we will continue to
use the total invariant mass
as an example in the following sections.

The weights in Eq. \ref{eq:scaling} are only for strangeness production, 
and they are relative to the production
weights of up and down quarks.
In the limit of a very heavy colour-singlet, the rate of producing
strangeness will be the same as that of the lighter quarks, while in the low-mass limit, only
the lighter quarks will be allowed to be produced.
The appeal of an exponential scaling is that this model only introduces one extra
parameter to the default model of hadronization in Herwig, and indeed, it does not introduce
any extra parameters if one splits the fission and decay parameters.
Thus we avoid a proliferation of parameters in our model,
and we still have a natural mechanism to allow for event-by-event fluctuations in 
strangeness production.

The scaling of the production rate in Eq. \ref{eq:scaling} only applies
to $s\bar{s}$ pairs, and not to any diquarks containing strange quarks. Default Herwig does
not allow gluons to non-perturbatively split into diquark-diantiquark pairs, nor does it allow
these pairs to be produced during cluster fissioning and decay.
Diquarks may only be produced as remnants of the incoming baryons, or from baryon-number
violating processes \cite{Herwig7}. Since diquark species would fundamentally affect the baryon
yields, which we are not studying in this work, we leave diquark production considerations to
a future rework of baryon production in Herwig.

\begin{figure*}[th]
\centering
\includegraphics[width=0.49\textwidth]{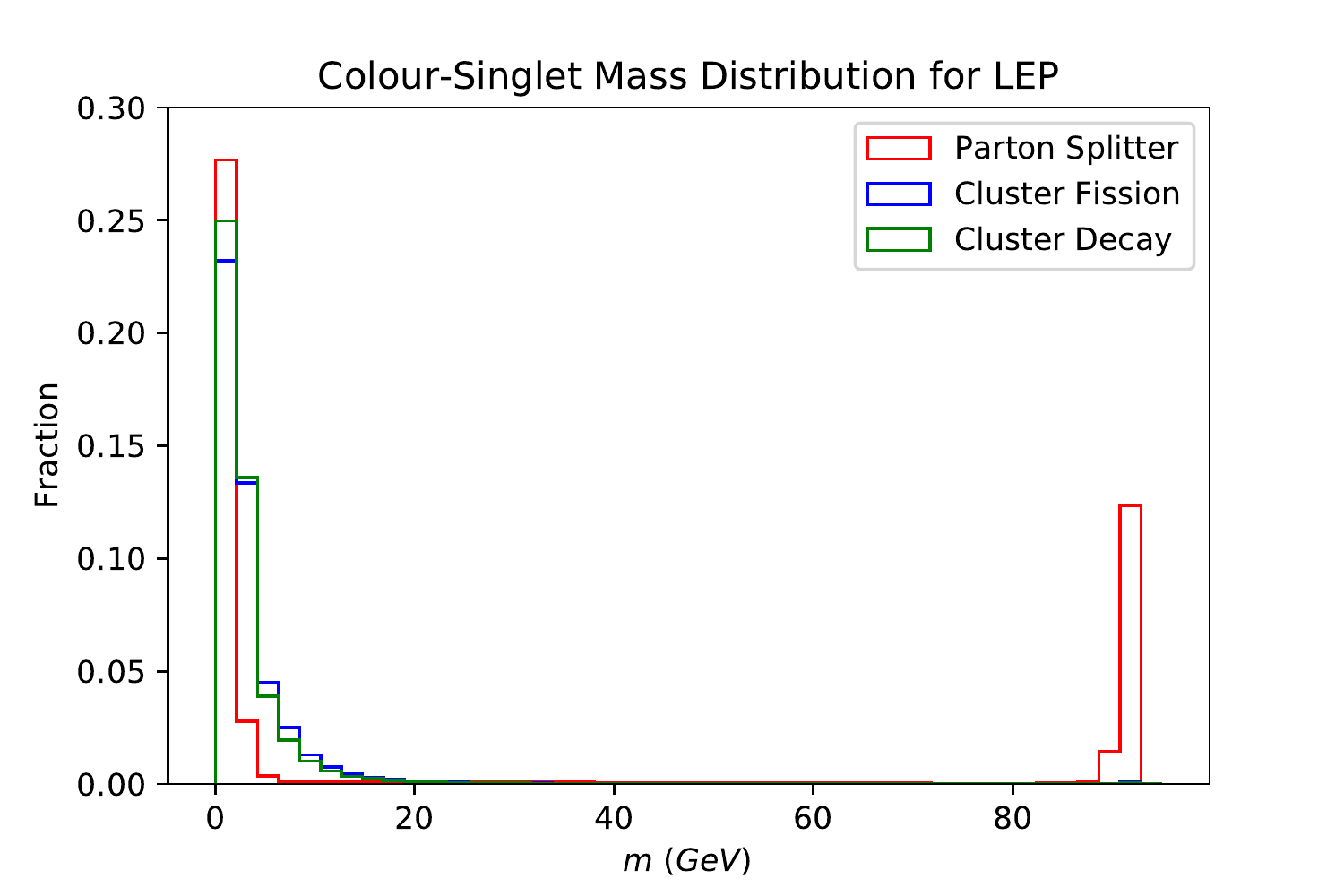}
\includegraphics[width=0.49\textwidth]{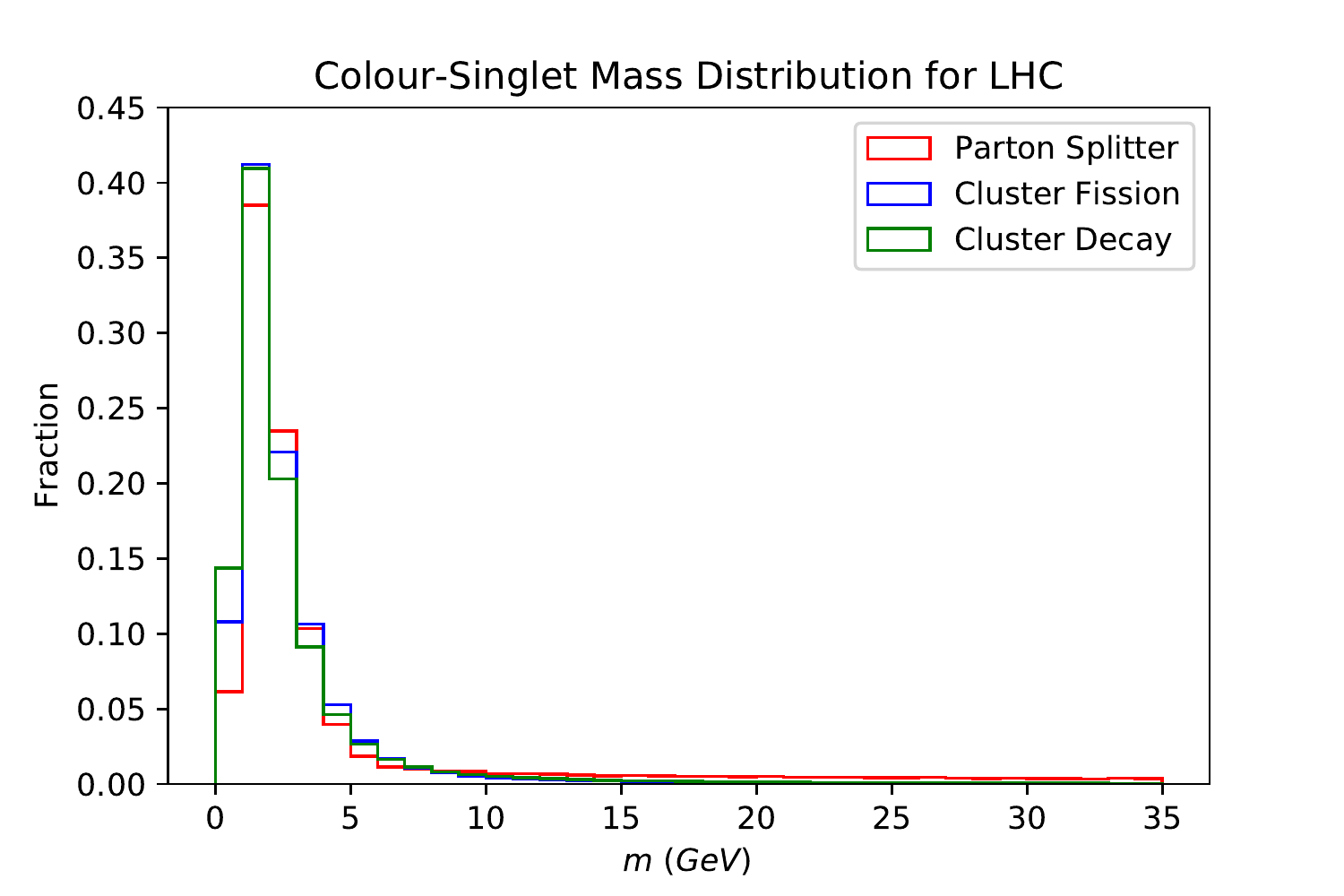}
\caption{Mass distributions for colour-singlet systems immediately 
before the Parton Splitter, Cluster Fissioner, and Cluster Decayer 
steps in LEP and LHC Minimum Bias events. Note the different mass axis scales.
}
\label{fig:massDist}
\end{figure*}

\begin{figure*}[th]
\centering
\includegraphics[width=0.49\textwidth]{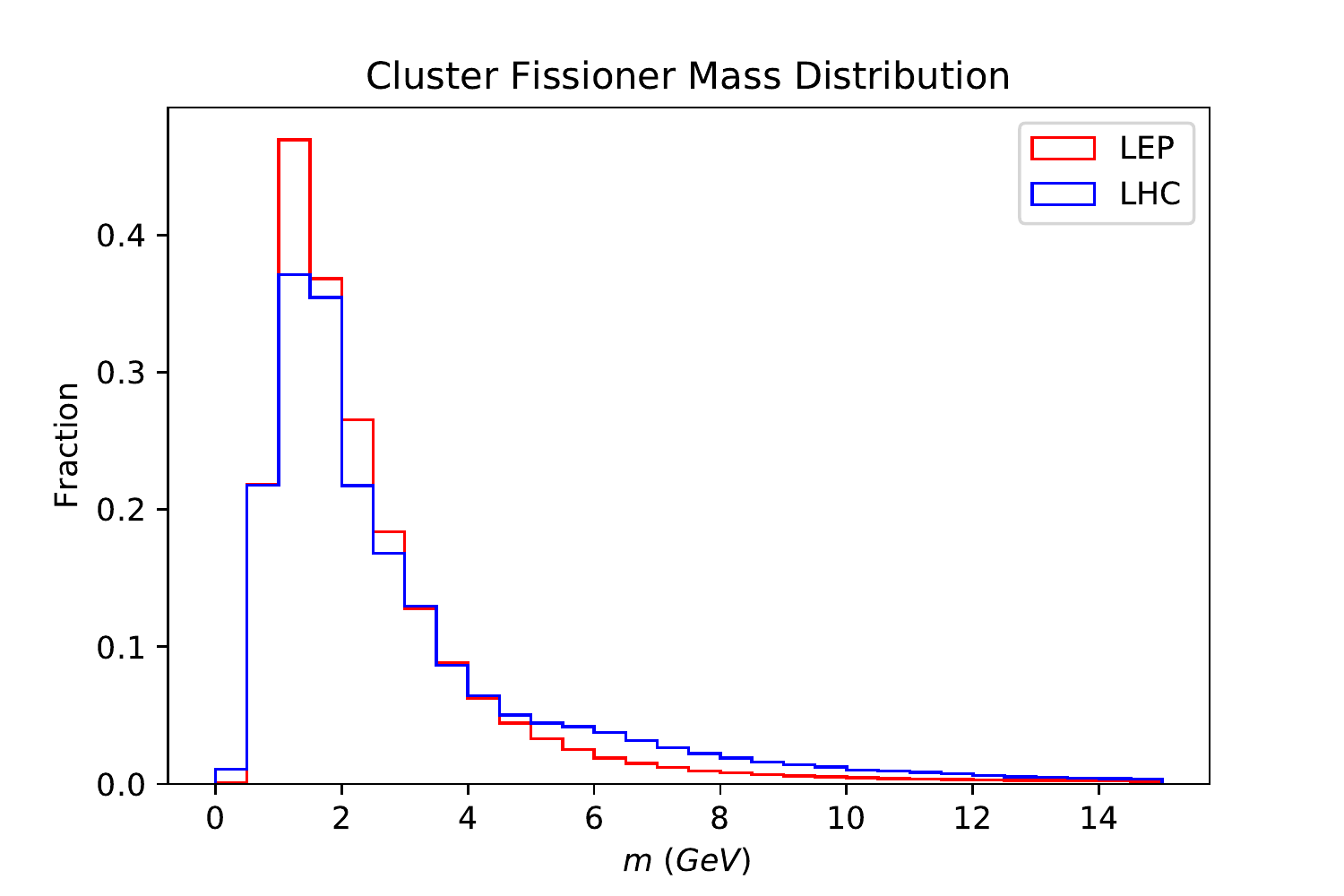}
\includegraphics[width=0.49\textwidth]{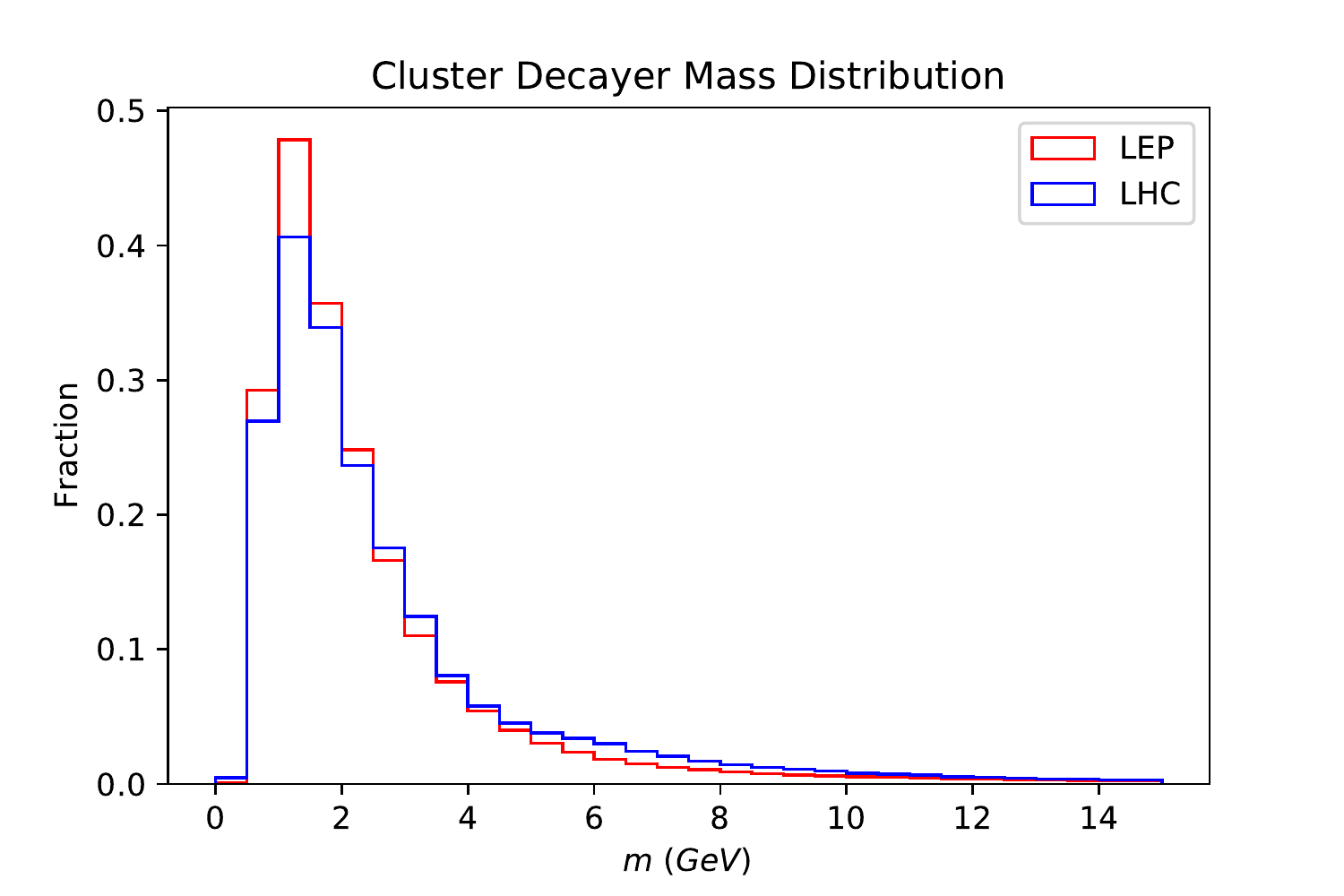}
\caption{Comparison of LEP and LHC Minimum Bias mass spectra of clusters 
immediately before cluster fission and cluster decay.
}
\label{fig:CFandCD}
\end{figure*}

\subsection{Non-perturbative gluon splitting}
\label{sec:newsplit}

\begin{figure}[t]
\centering
\includegraphics[width=0.49\textwidth]{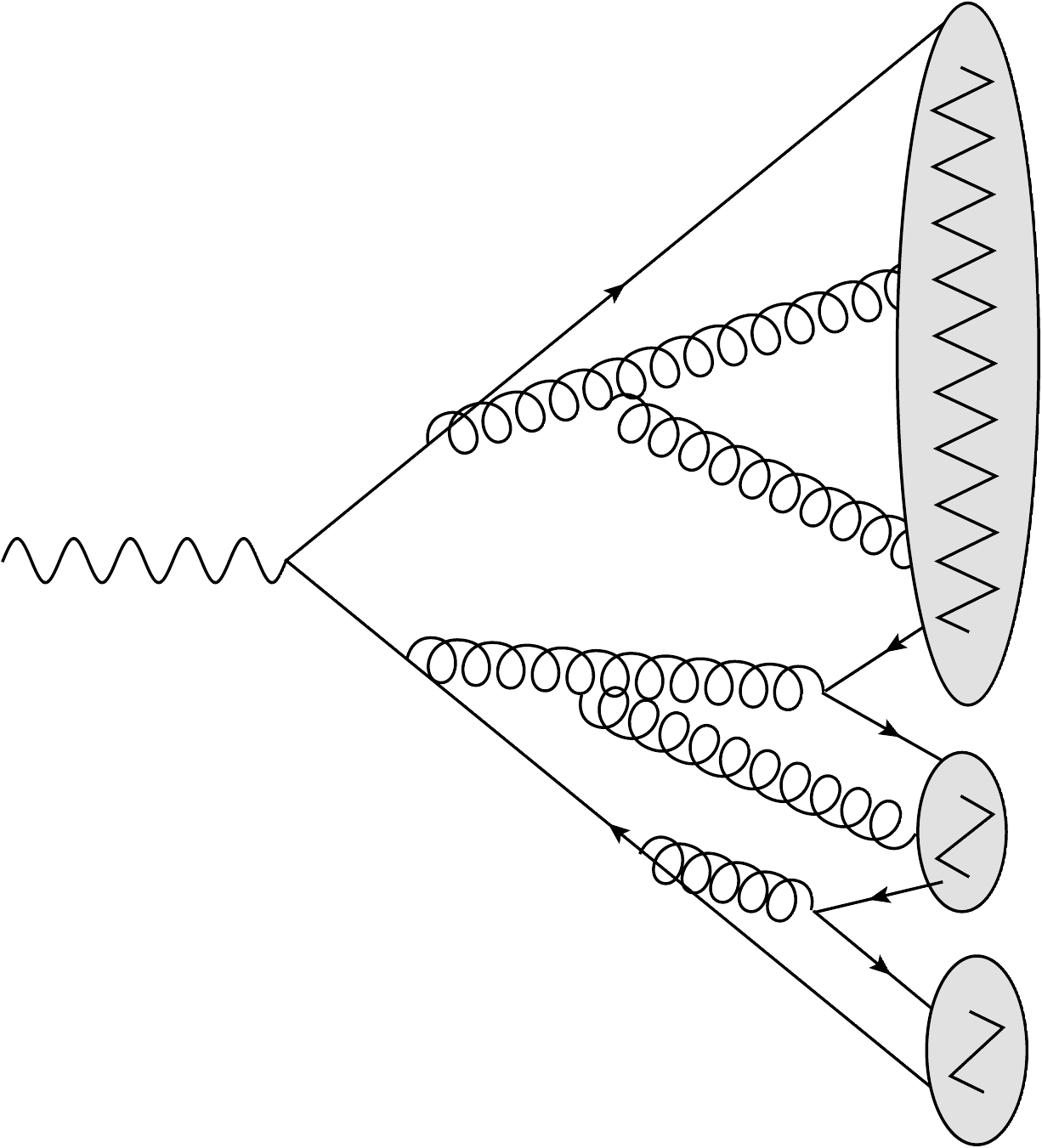}
\caption{Schematic topology of colour-singlets that can occur from perturbative gluon and quark shower
splitting, before the gluons undergo non-perturbative splitting.}
\label{fig:coloursinglet}
\end{figure}
At the end of the shower, instead of immediately splitting the gluons into $q\bar{q}$ pairs 
with the species determined by their given weights, we instead collect the various colour-singlet 
systems in the event, what we call \textit{pre-clusters}. 
While colour preconfinement dictates 
that the mass distribution of clusters is independent of the hard energy 
scale, there are no such constraints on the masses of the colour-singlet 
pre-clusters. As shown schematically in
Fig. \ref{fig:coloursinglet}, a parton shower can produce gluons and 
quark-antiquark pairs at a perturbative level, separating the event 
into a number of different pre-clusters with a variety of masses.

Every gluon in the same pre-cluster will get the same weight, since they belong
to the same colour-singlet system, and thus have the same mass measure
for strangeness production,
but since the species is picked probabilistically, this does not mean that all the gluons
will produce strange quark-antiquark pairs.
The constraint from default Herwig still applies, namely that even in situations where there
is a very heavy pre-cluster, if a gluon cannot access the phase space necessary 
to split into a $s\bar{s}$ pair, then it will undergo the usual splitting to up or 
down quarks.

The characteristic mass scale for pre-clusters will unfortunately depend on the
type of collider one uses. As shown in Fig. \ref{fig:massDist},
there is a very broad tail for the proton colliders due to the number of pre-clusters that
one can produce. This is a by-product of the type of dense and complicated
final state environment of high
energy hadron colliders. At LEP, there are two peaks for the pre-cluster mass distribution,
one at close to 91.2 GeV, corresponding to events where there are only gluon emissions
from the outgoing $q\bar{q}$ legs from the hard scattering process, and very few colour-singlets
fall between the two peaks, due to the simple fact that
perturbative gluon splitting is suppressed compared to perturbative gluon emission.

\subsection{Cluster fission \& decay}
At the cluster fission and cluster decay level, the colour-singlet is the cluster itself.
We allow the characteristic mass scale and characteristic production probability
to be different for the two phases. As shown in Fig. \ref{fig:CFandCD}, the typical cluster masses
at the cluster fission and cluster decay stages are roughly similar for both LEP and LHC,
which we hope to reflect in the characteristic mass scales for the two tunes.
We note that Figs. \ref{fig:massDist} and \ref{fig:CFandCD} are plotted without turning on
the exponential scaling, which would change the mass distribution slightly, but the figures
are benchmarks of the typical colour-singlet total invariant masses.

\subsection{Colour-singlet masses}
\label{sec:lambda}
In the previous sections we have used the total invariant mass of the colour-singlet systems
as the mass measure in Eq. \ref{eq:scaling}, but there are issues with this approach.
In using the total invariant mass of a given colour-singlet to scale the strangeness
weight, we have neglected to take into account the massive nature of the partons in
the pre-clusters and clusters. We argue that given two colour-singlets of the same total
invariant mass, if one cluster has much heavier endpoints or constituents that the other, 
then the one with lighter endpoints or constituents should more readily produce $s\bar{s}$
pairs from the vacuum.

To remove the biasing effects of massive constituents, we have implemented another
mass measure:
\begin{equation}
	\lambda = m_{cs}^2 - \left(\sum_i m_i\right)^2 ,
	\label{eq:lambda}
\end{equation}
where $m_{cs}^2$ is the total invariant mass of the colour-singlet system, and
$m_i$ are the invariant masses of the endpoints for pre-clusters or the constituent
partons in a cluster.

\begin{figure*}[t]
\centering
\includegraphics[width=0.49\textwidth]{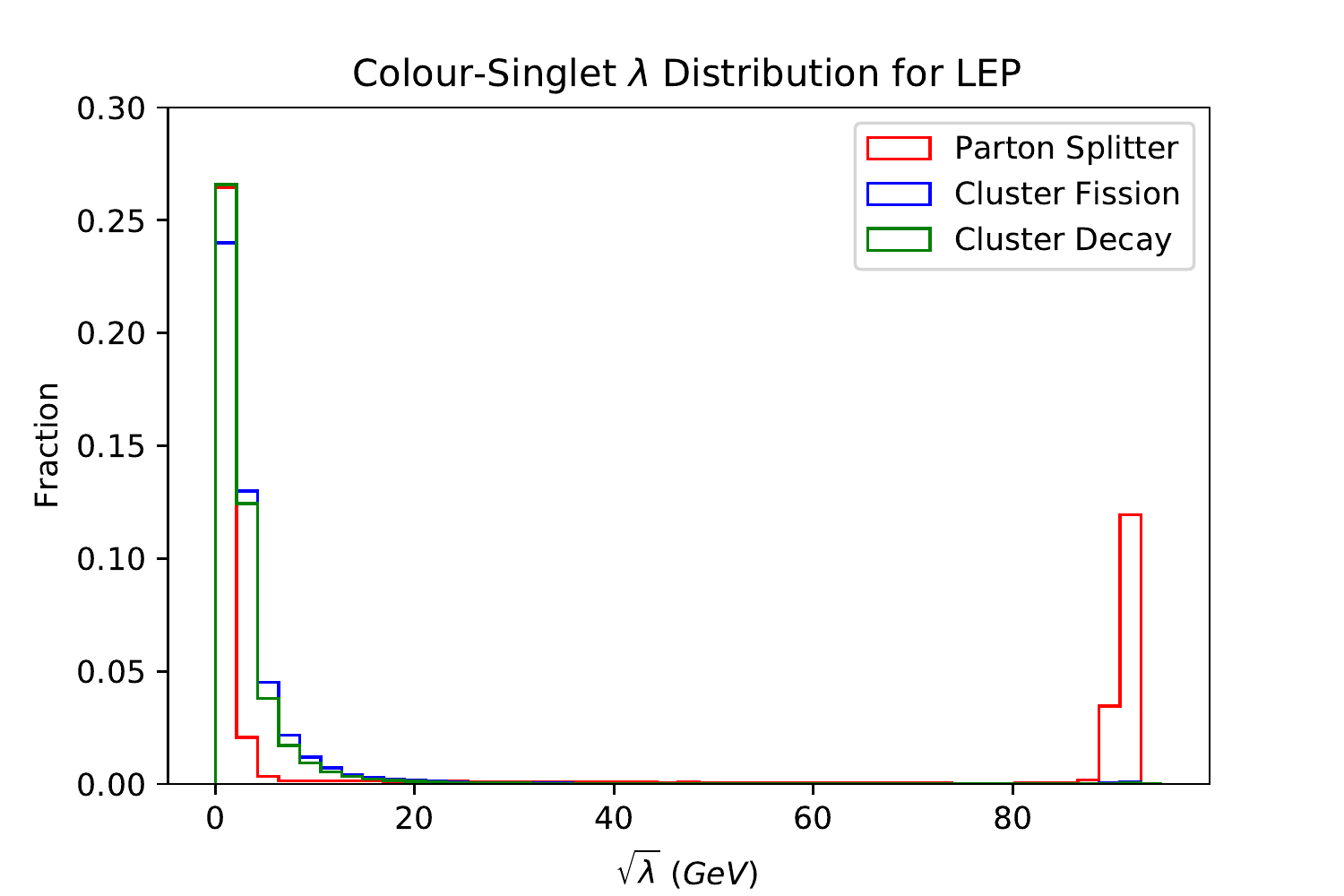}
\includegraphics[width=0.49\textwidth]{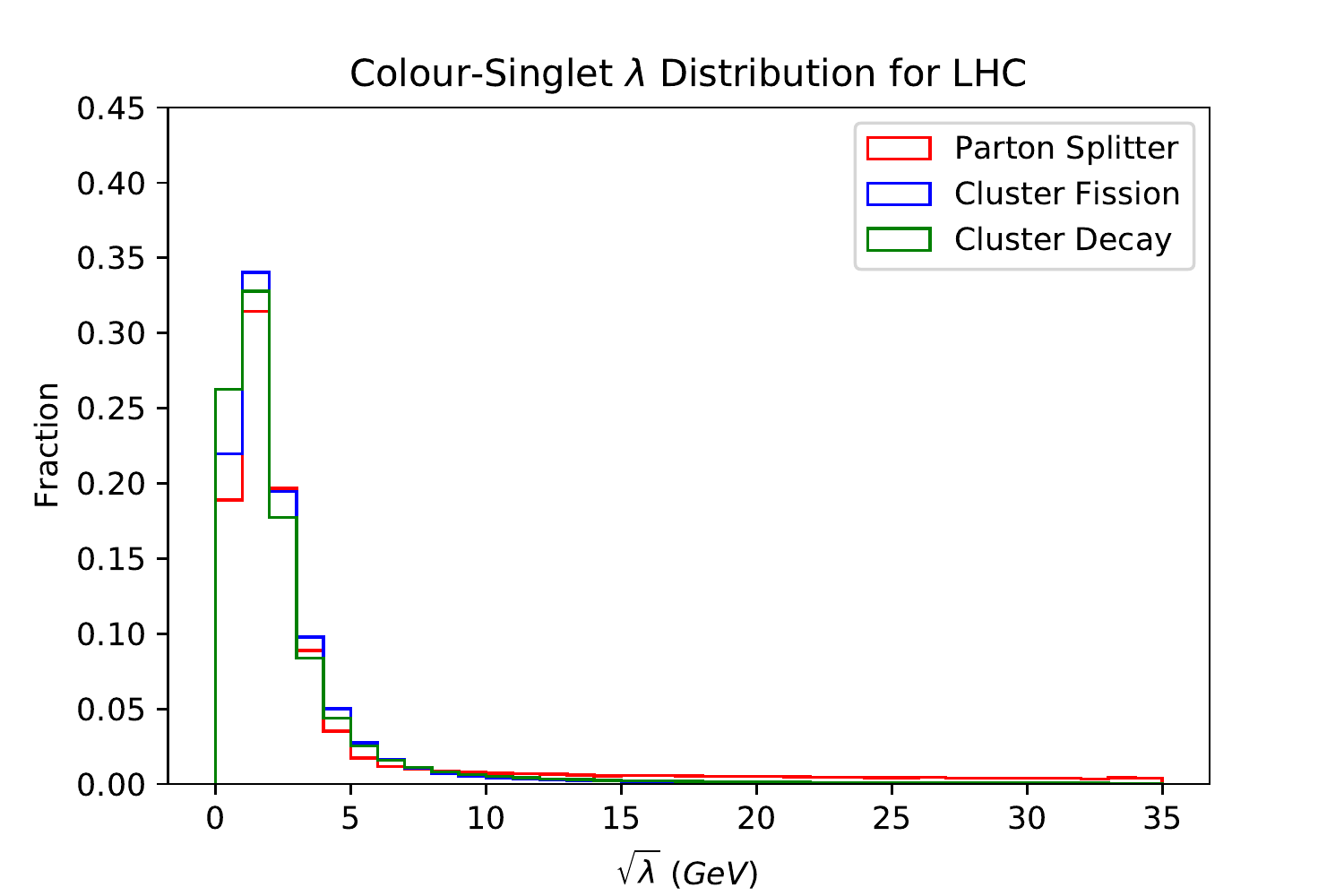}
\caption{Threshold mass, $\lambda$, distributions for colour-singlet systems 
immediately before the Parton Splitter, Cluster Fissioner, 
and Cluster Decayer steps at LEP events at 91.2 GeV and
LHC Minimum Bias events at 7 TeV.
}
\label{fig:lambdaDist}
\end{figure*}

\begin{figure*}[th]
\centering
\includegraphics[width=0.49\textwidth]{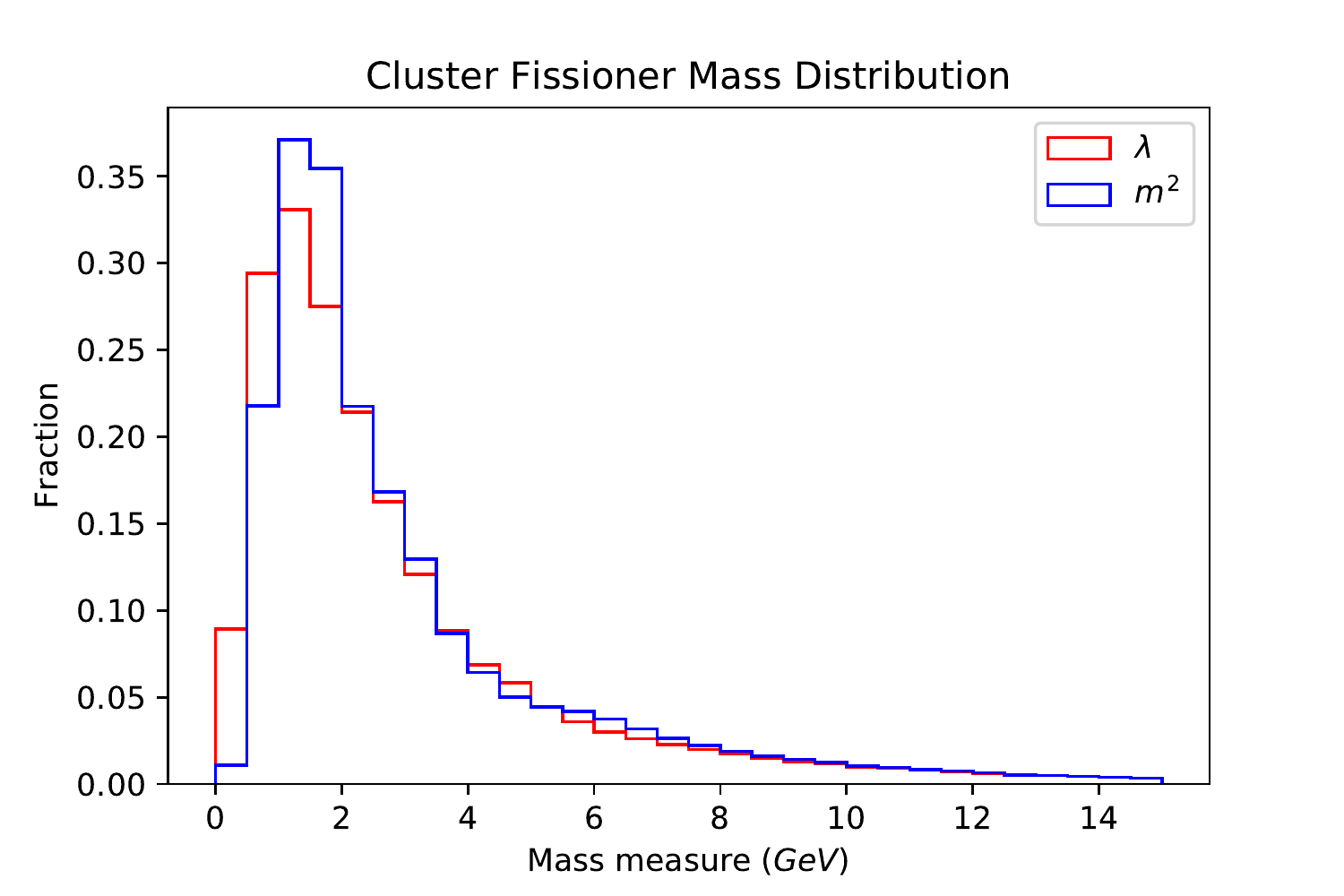}
\includegraphics[width=0.49\textwidth]{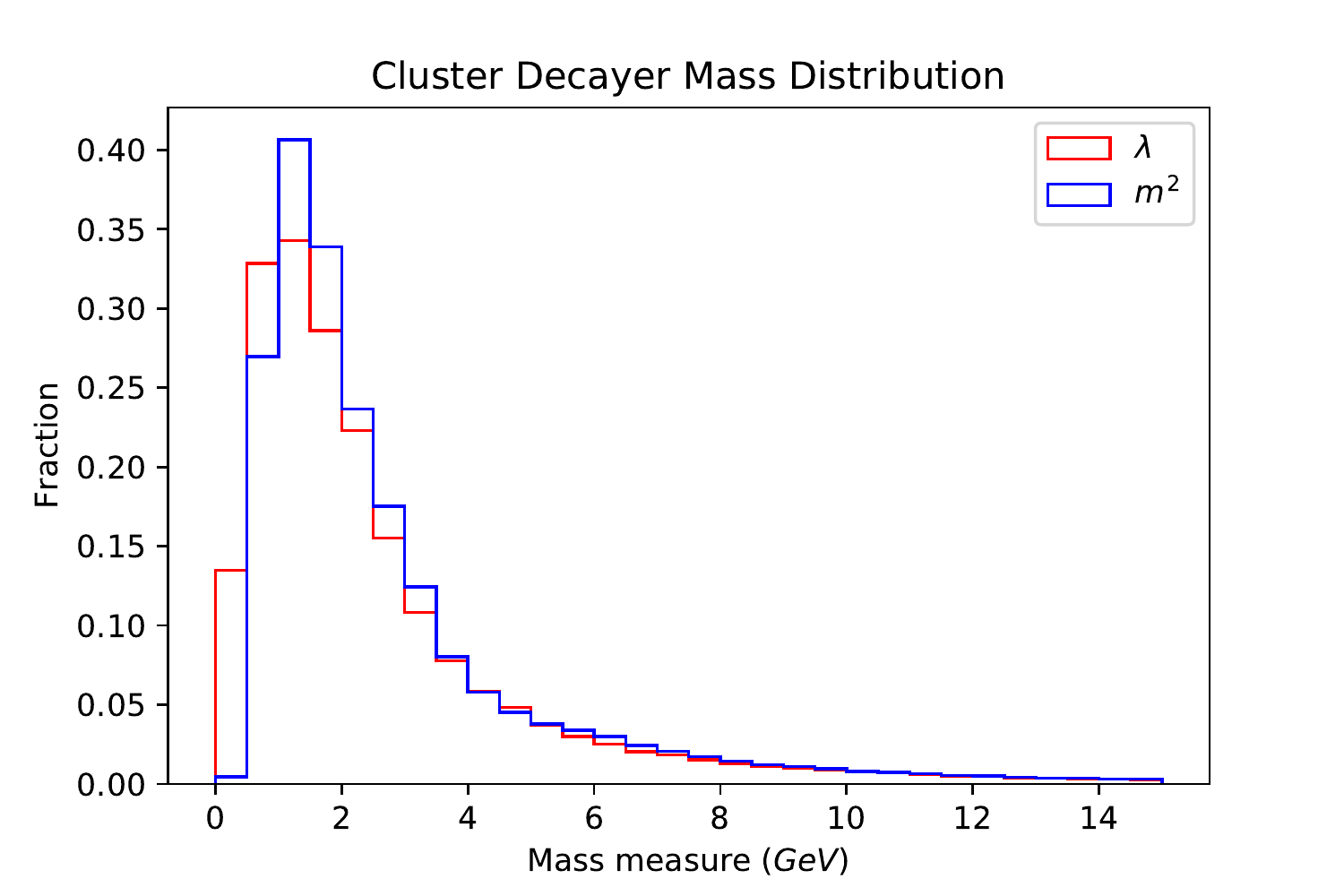}
\caption{Comparison of the two different mass measures for the cluster fission
and cluster decayer stages respectively for LEP events at 91.2 GeV.
}
\label{fig:mVlambdaLEP}
\end{figure*}

\begin{figure*}[th]
\centering
\includegraphics[width=0.49\textwidth]{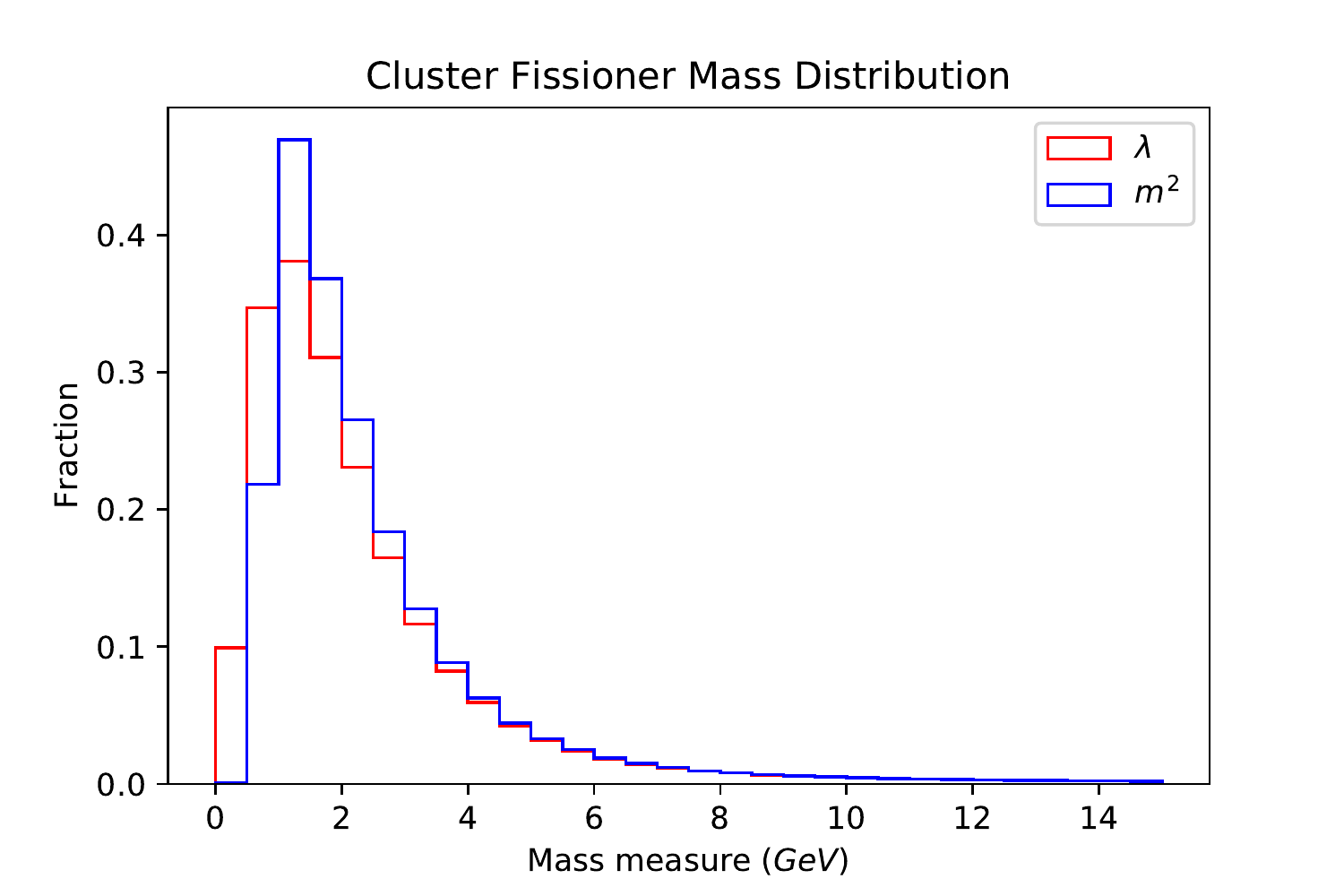}
\includegraphics[width=0.49\textwidth]{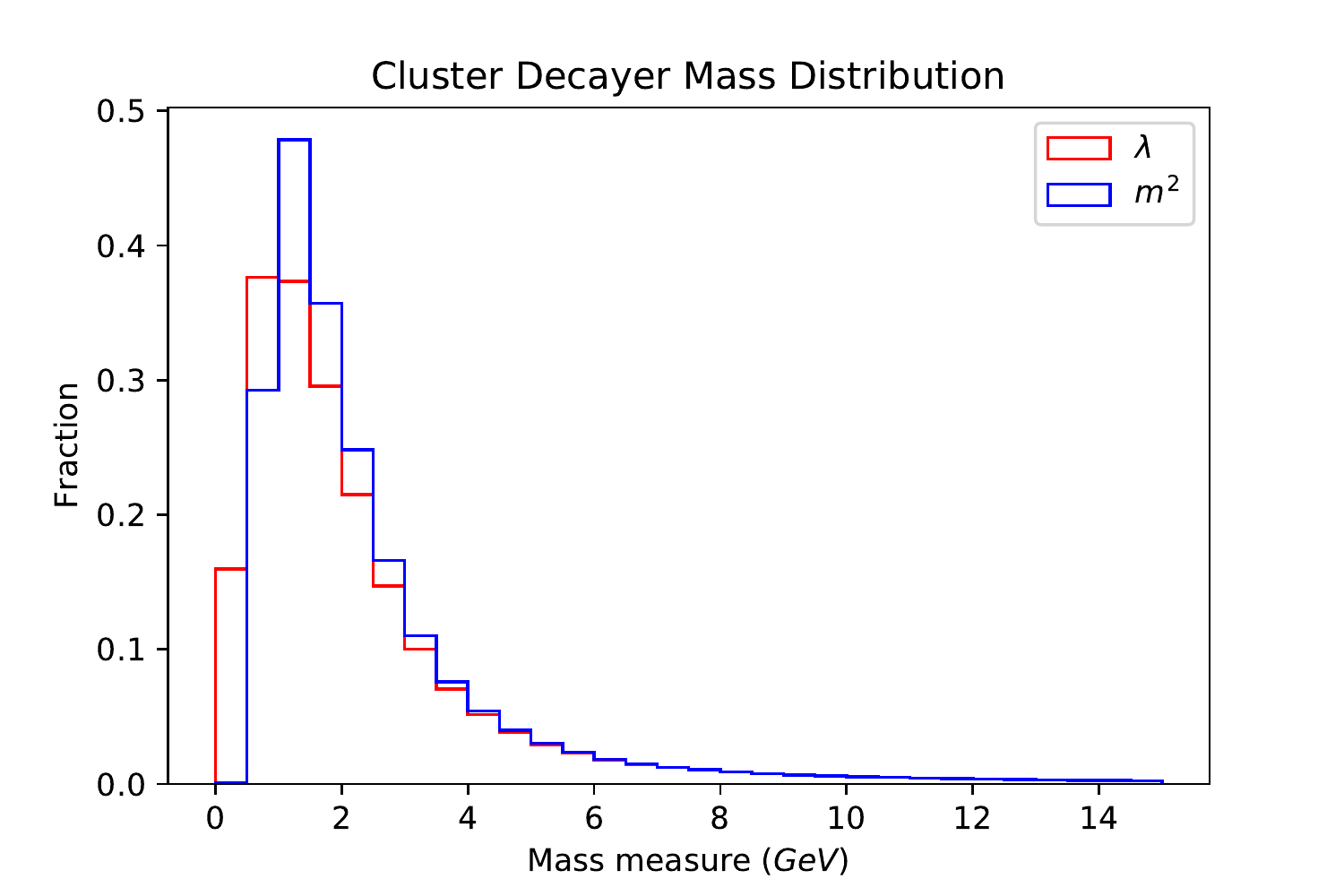}
\caption{Comparison of the two different mass measures for the cluster fission
and cluster decayer stages respectively for LHC Minimum Bias events at 7 TeV.
}
\label{fig:mVlambdaLHC}
\end{figure*}

Gluons are massive in Herwig, but because their masses are used
to produce the $s\bar{s}$ pair, we do not include them in the subtraction term.
The $\lambda$ measure would replace the mass-based denominator in Eq. \ref{eq:scaling}.
We have presented the distributions of the $\lambda$ measure for each of the stages 
in Fig. \ref{fig:lambdaDist}, and a comparison between
the distributions of the two mass measures in Figs. \ref{fig:mVlambdaLHC} and
\ref{fig:mVlambdaLEP}. 
The $\lambda$ measure has the appealing feature that if one produced a $s\bar{s}$ pair
at the gluon splitting level, this extra mass wouldn't propagate extra strangeness enhancement
further into the hadronization process.

\section{Analysis}
\label{sec:analysis}
We first tune the 3 parameters of our mass-based scaling model
to the same identified strange particle yields at LEP and LHC as in 
Sec.\, \ref{sec:oldtune}.
The new tunable parameters are \mbox{\tt MassScale} (for gluon splitting),
\mbox{\tt FissionMassScale}, and 
\mbox{\tt DecayMassScale}, which are defined by Eq.\,\ref{eq:scaling}.
The outcome of the tuning procedure for the relevant parameter values is
shown in Tabs.\,\ref{table:newmodelmass} and \ref{table:newmodellambda} 
for LEP and LHC Minimum Bias, for both the total invariant mass measure and
the $\lambda$ measure.

\begin{figure*}[th]
\centering
\includegraphics[width=0.49\textwidth]{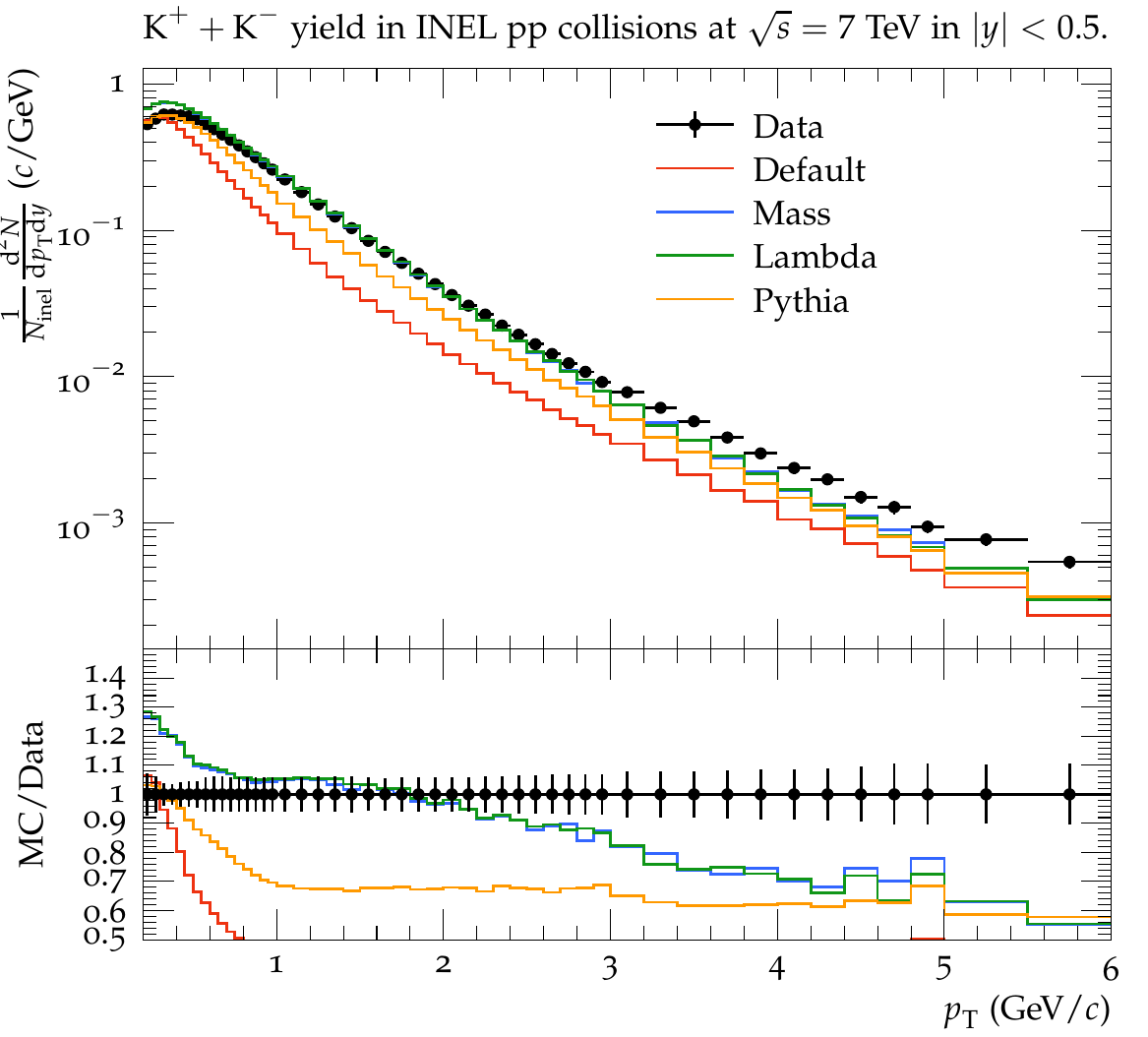}
\includegraphics[width=0.49\textwidth]{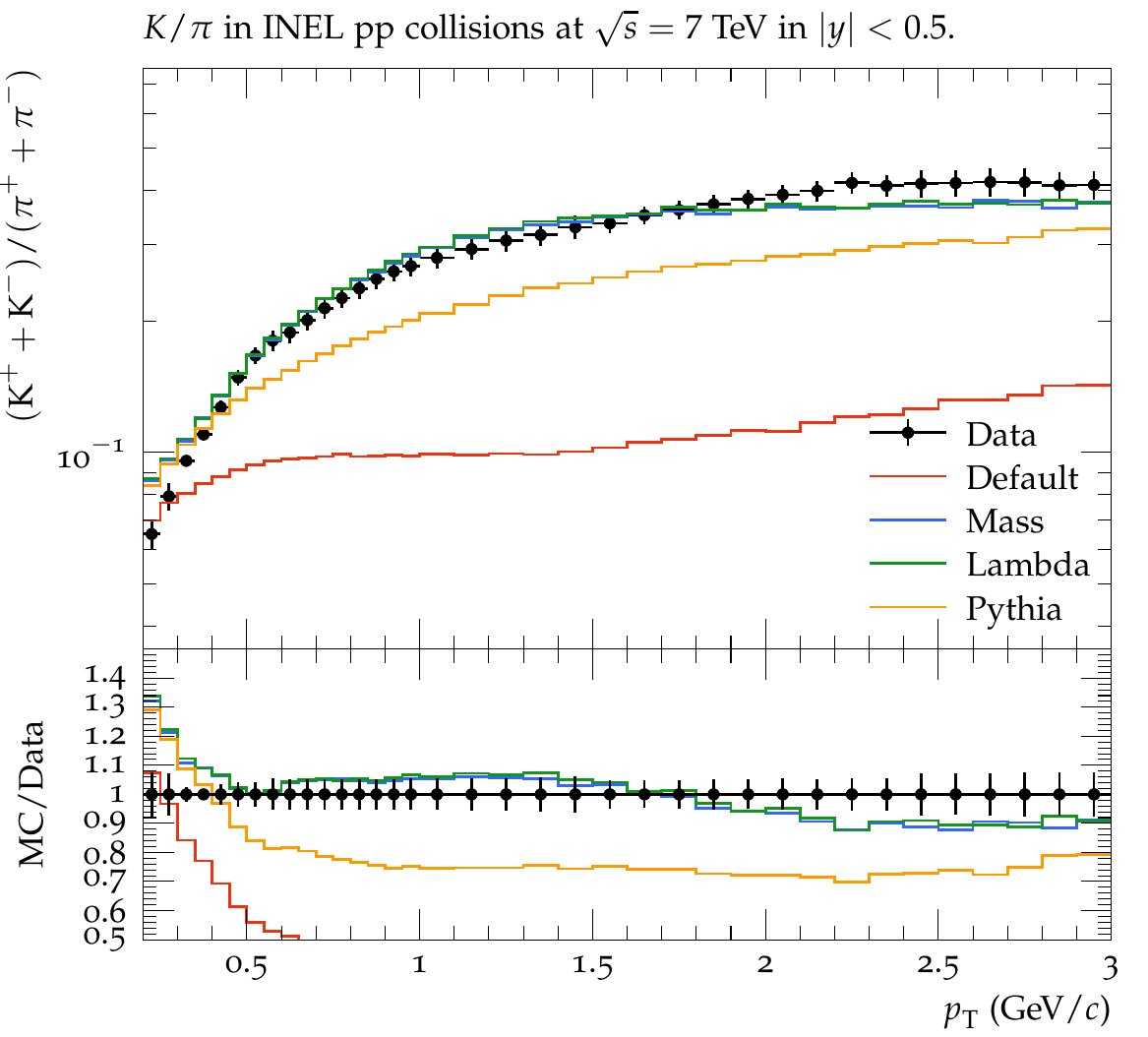}
\caption{$K^++K^-$ yield and $K/\pi$ ratio as measured by ALICE\cite{Adam:2015qaa} at 7 TeV.
Shown is a comparison between the default version of Herwig (without baryonic reconnection),
 i.e. static production of 
strangeness, the new approach which introduces dynamical strangeness production with
the two different measures (Mass and Lambda) and Pythia with the Monash tune.
} 
\label{fig:newmodeltuning}
\end{figure*}

\begin{table*}[htpb]
\parbox{0.45\textwidth}{
\centering
\begin{tabular}{lcc}
   \textbf{Invariant Mass} & LEP & LHC \\
\midrule
Gluon Splitting        & 97 & 48\\
Cluster Fission & 3  & 22\\
Cluster Decay   & 23 & 4\\
\end{tabular}
\caption {Results for the tuned characteristic mass scales $m_0$, 
in units of GeV, of our new model using the total 
invariant mass of a colour-singlet object for LEP and LHC tunes respectively.
  }
\label{table:newmodelmass}
}
\hfill
\parbox{0.45\textwidth}{
\centering
\begin{tabular}{lcc}
    $\boldsymbol{\lambda}$ \textbf{Measure} & LEP & LHC \\
\midrule
Gluon Splitting        & 72 & 37 \\
Cluster Fission        & 4 & 20 \\
Cluster Decay   & 16 & 10 \\
\end{tabular}
\caption {Results for the tuned characteristic mass scales $m_0$, 
in units of GeV, of our new model using our 
$\lambda$ measure (defined in Eq. \ref{eq:lambda}) of a colour-singlet 
object for LEP and LHC tunes respectively.
  }
\label{table:newmodellambda}
}
\end{table*}
With the three new characteristic mass scales, we are 
able to improve the description of all observables considered in the
tuning especially for LHC observables as shown in Fig.\ref{fig:newmodeltuning}, 
where we compare the two different mass measures after tuning,
as well as the Monash tune \cite{Skands:2014pea} for Pythia.

\begin{figure}[t]
\centering
\includegraphics[width=0.49\textwidth]{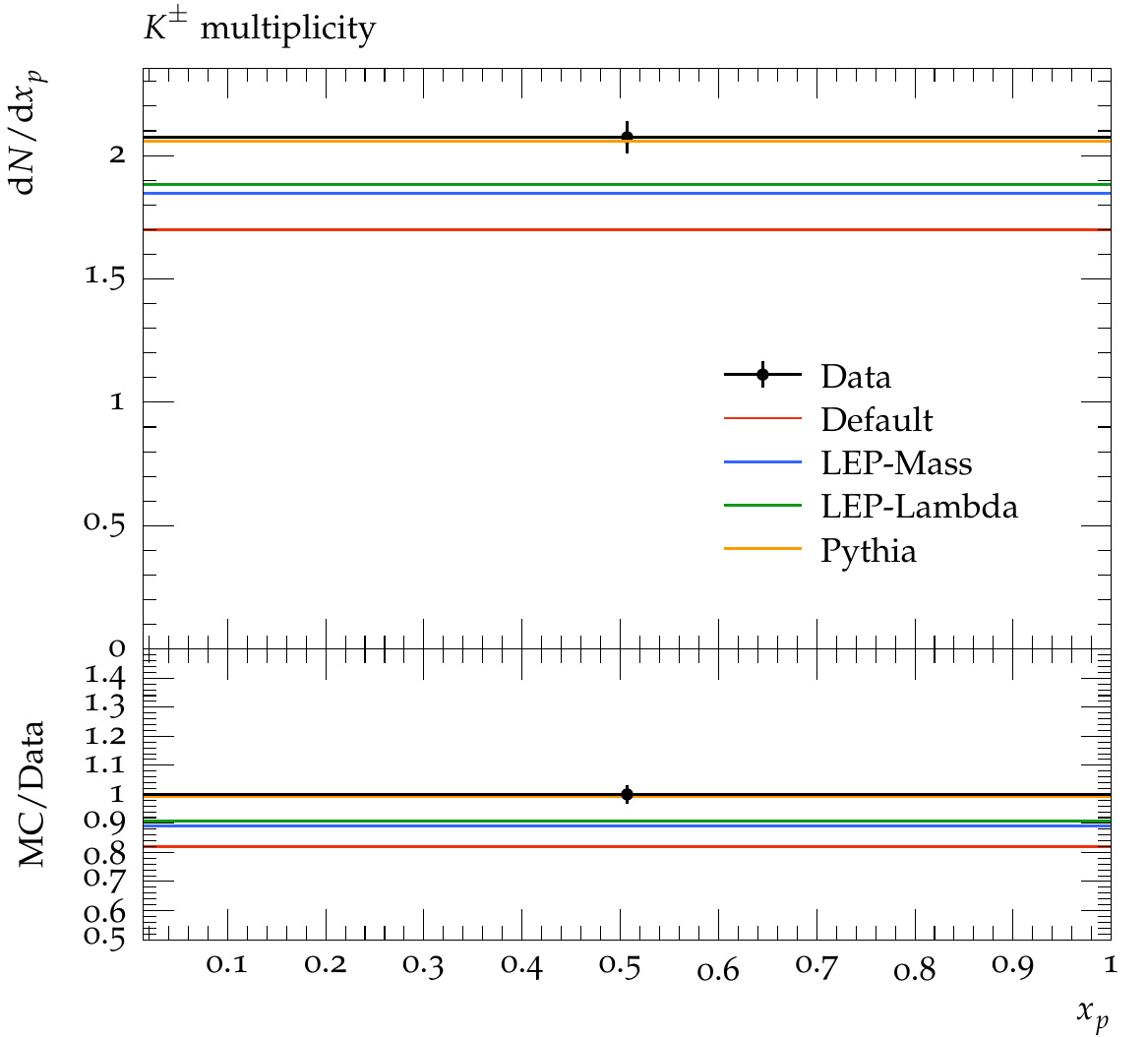}
\caption{ Measurement of $K^{\pm}$ multiplicities at SLD \cite{Abe:2003iy}
$\sqrt{s}=91.2\,\mathrm{GeV}$. We show a comparison between the default Herwig model
and the dynamical strangeness production where we used the LHC-tuned parameters (see Tabs. \ref{table:newmodelmass} and \ref{table:newmodellambda}) and Pythia with the Monash tune.
}
\label{fig:leptuningLHC}
\end{figure}

Although the simple tuning recommends different values for the usage at LHC and LEP
it is also feasible to use the set of parameters obtained from the tuning
to LHC data and still get improved results for LEP observables which was not possible 
by having a simple flat number as the probability to produce strange quarks as
is shown in Fig.\ref{fig:leptuningLHC}.

\subsection{Discussion}
The default version of Herwig did not allow for strange production during
the gluon splitting stage. By allowing this process, improvements can be seen 
in all the considered observables. With our new model, there is a more physically 
motivated dynamic strangeness production mechanism 
at all stages of the hadronization.

The multiple parton interaction model in Herwig involves two types of subprocesses,
hard and soft. Hard processes are allowed to shower and emit quarks and gluons,
while soft ones produce only gluons which may not shower. These soft gluons are all
colour-connected to each other and the beam remnants, resulting in a single
pre-cluster when undergoing non-perturbative gluon splitting. This type of pre-cluster
typically has a large invariant mass due to the large number of soft gluons and
the isotropic nature of their momentum distribution, resulting in a high strangeness production
weight for this subsystem. The resulting produced strange particles coming from these
soft interactions are distributed uniformly in rapidity.

There are three key differences between the LEP and LHC environments during 
hadronization. Firstly, LEP has a much lower energy scale than the LHC,
naturally limiting the possible distribution of colour-singlet masses at the stage of
non-perturbative gluon splittings. As a result, a direct comparison between
LEP and LHC in our model is not straightforward.

Secondly, while LEP and LHC simulations may have very similar cluster mass 
distributions, the number of clusters is far higher for the latter. Similarly, at the pre-cluster level,
LEP prefers colour-singlets that span the entire final state, as shown in 
Fig. \ref{fig:massDist}, i.e. no perturbative gluon splittings during the parton shower.
This results in the majority of events either having enhanced strangeness production
or none at all, at the gluon splitting level, meaning that a flat weight at this level
in hadronization can be justified for LEP runs.

Finally, and related to the previous two, LEP is a much cleaner environment.
For lepton collisions, there are no multiple parton interactions, 
nor much effect from colour reconnection. However, in proton collisions, these are both
vital phases of the simulation that drastically change the mass topology of the
event.

\begin{table*}[htpb]
\parbox{0.45\textwidth}{
\centering
\begin{tabular}{lcc}
   $\boldsymbol{E(w_s)}$ \textbf{at LEP}  & Mass & $\lambda$ \\
\midrule
Gluon Splitting  & 0.096 & 0.164\\
Cluster Fission  & 0.297  & 0.166\\
Cluster Decay   & 0.009 & 0.016\\
\end{tabular}
\caption {Expected value of strangeness production weight 
of our new model in LEP events at 91.2 GeV, comparing the total 
invariant mass results with the $\lambda$ measure results
  }
\label{table:convertedLEP}
}
\hfill
\parbox{0.45\textwidth}{
\centering
\begin{tabular}{lcc}
   $\boldsymbol{E(w_s)}$ \textbf{at LHC} & Mass & $\lambda$ \\
\midrule
Gluon Splitting  & 0.555 & 0.571\\
Cluster Fission  & 0.018  & 0.020\\
Cluster Decay   & 0.153 & 0.041\\
\end{tabular}
\caption {Expected value of strangeness production weight 
of our new model in LHC Minimum Bias events at 7 TeV, comparing the total 
invariant mass results with the $\lambda$ measure results
  }
\label{table:convertedLHC}
}
\end{table*}

Taking the characteristic mass scales from Tabs. \ref{table:newmodelmass}
 and \ref{table:newmodellambda}, we have translated these into an effective
expected value for the weights for the two mass measures.
For LEP events, as shown in Tab. \ref{table:convertedLEP}, the total invariant mass approach
prefers cluster fissioning, while for the $\lambda$ measure, non-perturbative
gluon splitting and cluster fissioning are approximately the same.
It should be noted that aside from the gluon splitting weights, there is
no direct translation between the kinematic picture and the old model of
strangeness production, but these expected values give an idea of the
average weights. For gluon splitting at LEP, the weight simply varies between 0 and
the maximal value, since pre-clusters are predominately situated around two peaks, as
shown in Fig. \ref{fig:massDist}, and the value shown in Tab. \ref{table:convertedLEP}
is simply half the maximal value of 0.192 in the invariant mass case, and 0.328 for the $\lambda$
measure.

For LHC Minimum Bias events, the expected value for the weights are shown in Tab.
\ref{table:convertedLHC}. There is very little difference between using the two mass
measures at the gluon splitting and cluster fission stages, while cluster decay is significantly
suppressed when using the $\lambda$ measure. The enormous suppression of 
strangeness production during the later stages of hadronization
compared to the gluon splitting is almost certainly a
hint that colour reconnection plays a non-trivial role in producing strange hadrons.
Our new kinematic model uses a mass-based scaling, but colour reconnection aims
to lower the cluster masses to some local minimum, meaning that it is in direct conflict with
our considerations. For LEP simulations, colour reconnection has a 
small effect, while in LHC simulations, colour reconnection is a vital phenomenon.
Future work will study the correlations between the role colour reconnection
plays and our model, in particular, varying the amount of colour reconnection that takes
place in an event, and allowing baryonic clusters to form.

Our studies showed that there is virtually no quantitative difference between
using the tuned invariant mass parameters and the tuned $\lambda$ measure parameters.
However, the results in Tabs.
\ref{table:convertedLEP} and \ref{table:convertedLHC} suggest that the $\lambda$
measure bridges the divide between the two types of collision better.

We have also compared the results of our new model with Pythia and the Monash tune
in Figs. \ref{fig:newmodeltuning} and \ref{fig:leptuningLHC}. While the Monash tune
aims to describe a number of observables other than the strangeness production
rate in Pythia, it is tuned to both LEP and LHC data \cite{Skands:2014pea}, making it
an apt benchmark for this discussion.

\begin{figure*}[th]
\centering
\includegraphics[width=0.49\textwidth]{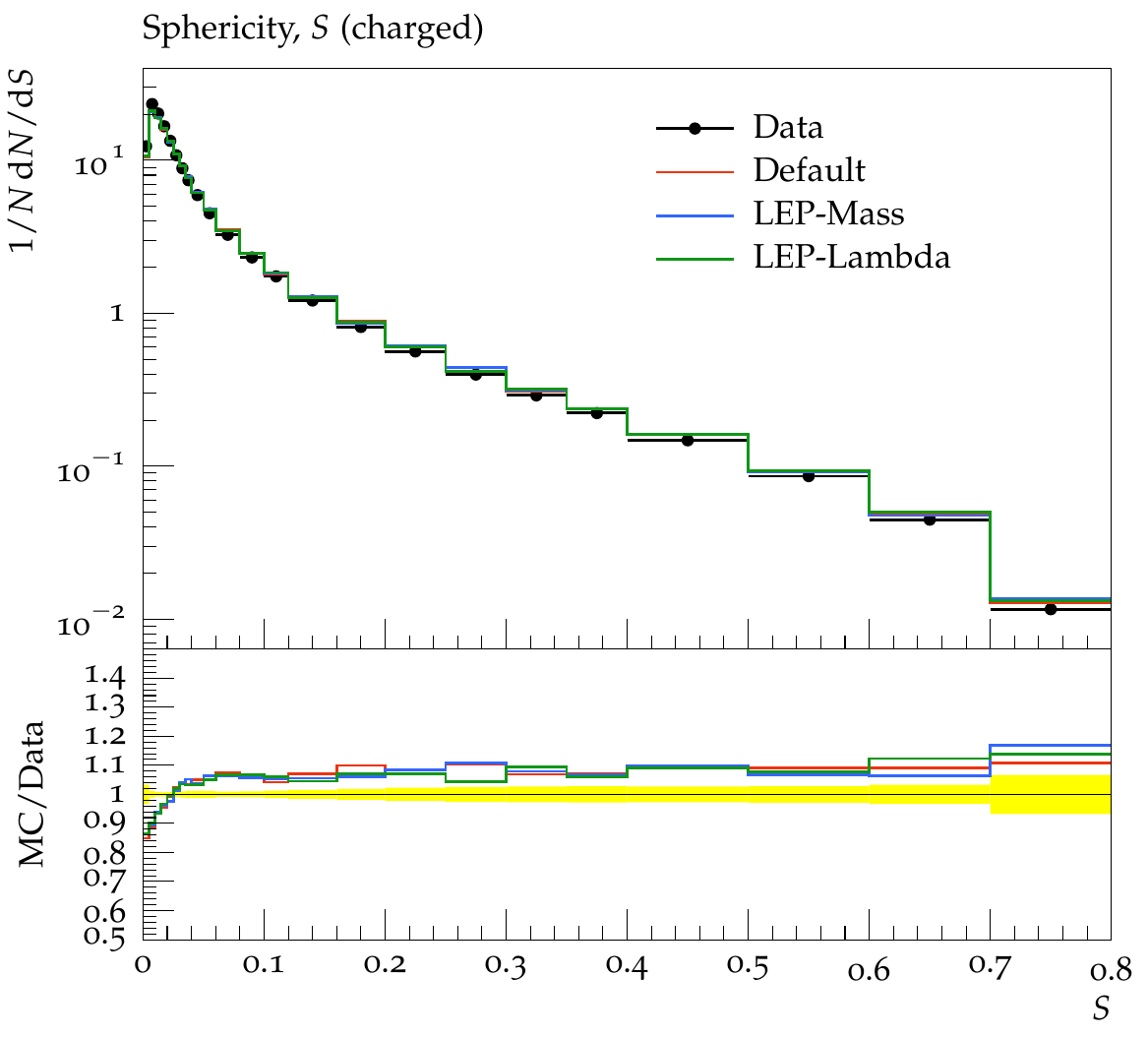}
\includegraphics[width=0.49\textwidth]{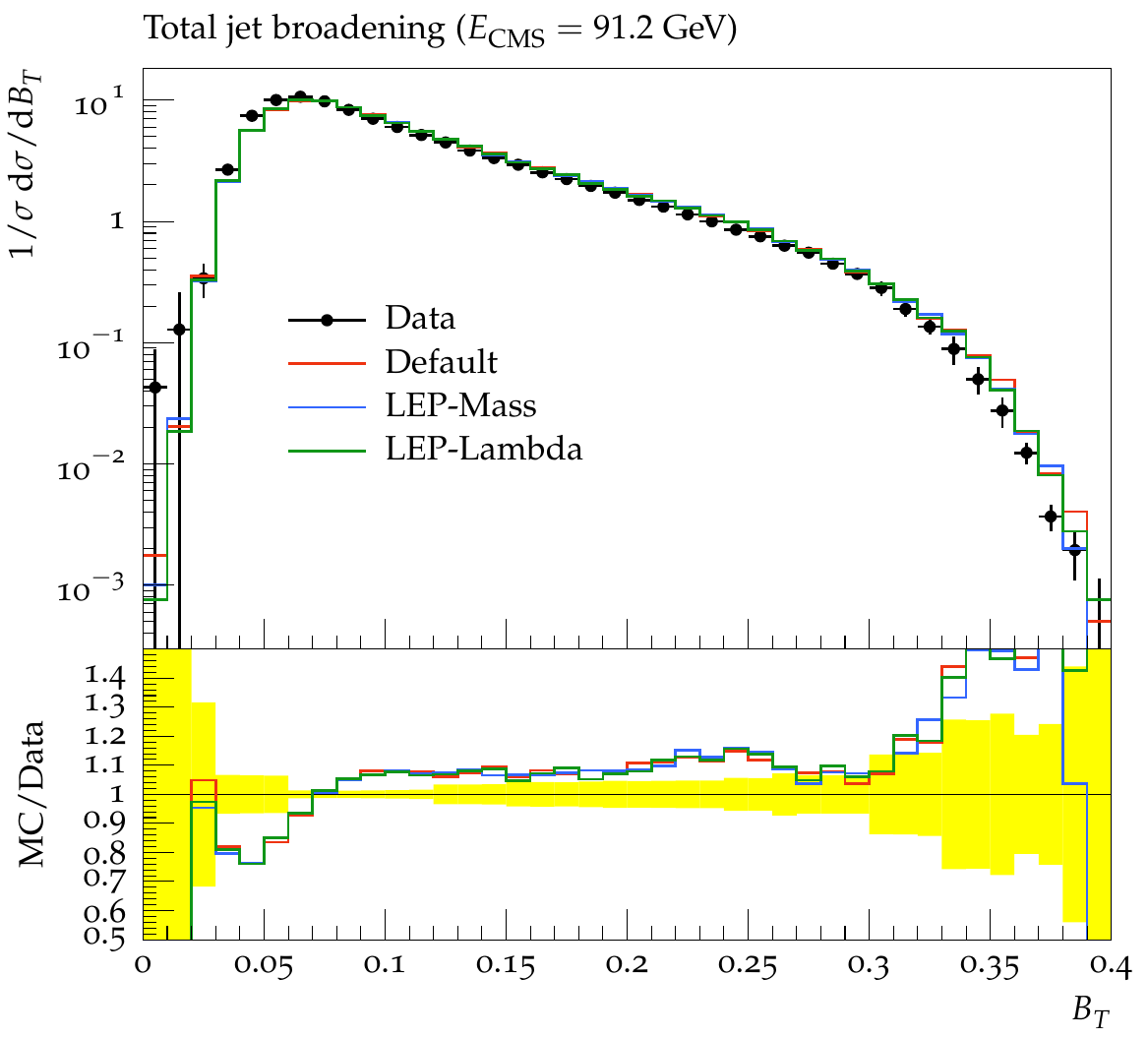}
\caption{Event-shape observables from ALEPH \cite{Barate:1996fi,
TheALEPHCollaboration2004}, comparing the results
of default Herwig to our new LEP tuned non-perturbative strangeness production scaling, for 
both mass and $\lambda$ measures. The new scaling does not impact on 
event-shape observables.
} 
\label{fig:eventshapes}
\end{figure*}

We can see that our model performs marginally better than Pythia, 
and significantly better than default Herwig, when trying to
describe the $K^{\pm}$ and drastically better on both counts for the $K/\pi$ ratio yields, 
as shown in Fig. \ref{fig:newmodeltuning}. However, in the low-$p_{\perp}$ region, both 
Pythia and our model overestimate the data. 
When using LHC Minimum Bias tuned parameters for LEP simulations, our model
outperforms the default Herwig model, but Pythia describes the data better, as shown
in Fig. \ref{fig:leptuningLHC}.

We expect that changing non-perturbative strangeness production scaling should
not change the overall event-shape observables, such as the Sphericity, 
and total jet broadening.
We have included several of these observables from ALEPH data \cite{Barate:1996fi,
TheALEPHCollaboration2004}
in Fig. \ref{fig:eventshapes}, 
to confirm that there are only minor statistical differences between default Herwig 7
and our new scaling when one is concerned with non-species specific observables.

While we have not fully solved the discrepancy between the weights for LEP and LHC
strangeness production, we have achieved two results: firstly, we have narrowed
the gap between the weights of the two types of collision, and in particular, our model
can be used with LHC Minimum Bias tuned parameters to better describe LEP data.
Secondly, we have made the first steps
to a more sophisticated treatment of hadronization and pair production
at the low-energy scale in Herwig.

\section{Conclusion and Outlook}
\label{sec:conclusion}
We have introduced a three-part model that scales the probability for strangeness
production during the hadronization phase of event generation in Herwig. The scaling is
directly controlled by the mass of the corresponding event colour-singlet 
subsystem at each step. With this mechanism, we allow for greater 
fluctuations in the production of strange pairs
on an event-by-event basis.

We have studied the mechanism for non-perturbative strangeness production in detail
and found that the current flat probability model is irreconcilable with both
LEP and LHC data. A hadronization model should be able to have minimal effects
on LEP simulations, but produce significant effects for LHC simulations.

After allowing a mass-based scaling, and tuning the parameters to LEP and LHC
data, we find that we are able to narrow the gap between the two collider 
types, and able to describe some observables better than the Lund string model
in Pythia with the Monash tune. We also provide expected values for non-perturbative
strangeness production, which capture the average values for event-by-event
fluctuations.

It should be noted that we have not considered heavier hyperons, the production 
of which has been shown to be increased by creating baryonic clusters at the 
colour reconnection stage \cite{Gieseke:2017clv}. Baryonic clusters, which are 
heavier by nature, would modify our model's strangeness production rates. 
Understanding the interplay between our new model and colour reconnection
will be left for future work.

There is still much left to understand in soft physics, but understanding the correlations
created between the various models in hadronization are imperative to having more precise
and useful Monte Carlo event generators.

\section*{Acknowledgements}
We would like to thank Stefan Gieseke and Peter Skands for their 
comments on the manuscript.
We would also like to thank Stefan Prestel for his helpful comments on the work and
Peter Christiansen for his invaluable help understanding experimental analyses.
This work has received funding from the European
Union’s Horizon 2020 research and innovation programme as part of the Marie 
Sklodowska-Curie Innovative Training Network MCnetITN3 (grant agreement no. 722104).
This work has been supported by the BMBF under grant number 05H18VKCC1.

\bibliography{main}

\end{document}